\documentclass[iop]{emulateapj}
\usepackage{amsmath}
\usepackage{amssymb}
\usepackage{epsfig}
\usepackage{subfigure}
\usepackage{natbib}
\usepackage{graphicx}
\usepackage{color}
\usepackage{longtable}
\usepackage{verbatim}

\shorttitle{PS1 SN IIP Light Curves}
\shortauthors{Sanders et al.}

\newcommand{\AndersonbetatwoMpCscatter}{0.3}

\newcommand{\PSOzeroonezeroonesixthree}{PS1-10ae}

\newcommand{\PSOzerosixoneoneninesix}{PS1-10zu}

\newcommand{\PSOonefivezerosixninetwo}{PS1-11wj}

\newcommand{\PSOthreezerozerotwotwoone}{PS1-11apd}

\newcommand{\PSOthreesevenzerothreethreezero}{PS1-12sz}

\newcommand{\PSOfourtwozerothreeninethree}{PS1-12bku}

\newcommand{\NIItotal}{112} 
\newcommand{\NIIPtotal}{67} 
\newcommand{\NIIbtotal}{5} 
\newcommand{\NIIntotal}{23} 
\newcommand{\NIIqtotal}{17} 
\newcommand{\NIIPtotalF}{76}

\newcommand{\NphotTot}{18,953} 
\newcommand{\NphotTotDet}{5,096} 
\newcommand{\SVMEffAll}{83} 
\newcommand{\SVMEffIIp}{89}

\newcommand{\SVMFPIIp}{22}

\newcommand{\SVMFNIIp}{11} 
\newcommand{\SVMClP}{9} 
\newcommand{\SVMCln}{3} 
\newcommand{\SVMClb}{6} 
\newcommand{\zdistIIAd}{0.07} 
\newcommand{\zdistIIAm}{0.11} 
\newcommand{\zdistIIAu}{0.21} 
\newcommand{\zdistIIPd}{0.07} 
\newcommand{\zdistIIPm}{0.10} 
\newcommand{\zdistIIPu}{0.16} 
\newcommand{\BETAkdefactor}{0.43} 
 
\newcommand{\betatwoMpPp}{1\times10^{-05}}

\newcommand{\betatwoMpBFVmed}{0.26}

\newcommand{\betatwoMpCP}{-0.50} 
\newcommand{\betatwoMpCPp}{5\times10^{-06}} 
\newcommand{\betatwoMpCBFamed}{-13.1} 
\newcommand{\betatwoMpCBFastd}{1.2} 
\newcommand{\betatwoMpCBFbmed}{-0.47} 
\newcommand{\betatwoMpCBFbstd}{0.07} 
\newcommand{\betatwoMpCBFVmed}{0.16} 
\newcommand{\betatwoMpCBFVdown}{0.14} 
\newcommand{\betatwoMpCBFVup}{0.19} 
\newcommand{\betatwoMpCscatter}{0.2} 
\newcommand{\stackNacceptg}{63} 
\newcommand{\stackNtemptotalg}{434} 
\newcommand{\stackNacceptr}{66} 
\newcommand{\stackNtemptotalr}{586} 
\newcommand{\stackNaccepti}{64} 
\newcommand{\stackNtemptotali}{760} 
\newcommand{\stackNacceptz}{63} 
\newcommand{\stackNtemptotalz}{757} 
\newcommand{\stackNaccepty}{32} 
\newcommand{\stackNtemptotaly}{267} 
\newcommand{\stackNtemptotalALL}{2,804} 
 
\newcommand{\MNiD}{0.04} 
\newcommand{\MNiM}{0.12} 
\newcommand{\MNiU}{0.20} 
\newcommand{\MNidUdex}{0.4} 
 
\newcommand{\tpfN}{29} 
 
\newcommand{\tpfM}{92} 
 
\newcommand{\tpfmin}{42} 
\newcommand{\tpfmax}{126} 
\newcommand{\tpfrstd}{14} 
\newcommand{\tpfrstdprior}{24} 
\newcommand{\EBVmed}{1.3}

\newcommand{\EBVmin}{0.0} 
\newcommand{\EBVmax}{4.8} 
\newcommand{\EBVsigmed}{0.4} 
 
\newcommand{\EBVsigNexc}{8} 
\newcommand{\mpeakNg}{36} 
\newcommand{\mpeakDg}{-18.43} 
\newcommand{\mpeakMg}{-17.86} 
\newcommand{\mpeakUg}{-16.86} 
 
\newcommand{\mpeakCNg}{13} 
\newcommand{\mpeakCDg}{-19.70} 
\newcommand{\mpeakCMg}{-19.14} 
\newcommand{\mpeakCUg}{-17.84} 
 
\newcommand{\mpeakNr}{50} 
\newcommand{\mpeakDr}{-18.45} 
\newcommand{\mpeakMr}{-17.80} 
\newcommand{\mpeakUr}{-16.92} 
\newcommand{\mpeakdmedr}{0.07} 
\newcommand{\mpeakCNr}{23} 
\newcommand{\mpeakCDr}{-19.28} 
\newcommand{\mpeakCMr}{-18.10} 
\newcommand{\mpeakCUr}{-17.47} 
\newcommand{\mpeakCdmedr}{0.35} 
\newcommand{\mpeakNi}{49} 
\newcommand{\mpeakDi}{-18.33} 
\newcommand{\mpeakMi}{-17.68} 
\newcommand{\mpeakUi}{-17.08} 
 
\newcommand{\mpeakCNi}{26} 
\newcommand{\mpeakCDi}{-18.94} 
\newcommand{\mpeakCMi}{-18.10} 
\newcommand{\mpeakCUi}{-17.31} 
 
\newcommand{\mpeakNz}{40} 
\newcommand{\mpeakDz}{-18.42} 
\newcommand{\mpeakMz}{-17.89} 
\newcommand{\mpeakUz}{-17.11} 
 
\newcommand{\mpeakCNz}{33} 
\newcommand{\mpeakCDz}{-19.25} 
\newcommand{\mpeakCMz}{-18.06} 
\newcommand{\mpeakCUz}{-17.63}

\newcommand{\mpeakCNy}{15}

\newcommand{\MinNoutofbounds}{67} 
\newcommand{\MinNoutofboundslow}{61} 
\newcommand{\MinNinbounds}{8}

\newcommand{\gps}{\ensuremath{g_{\rm P1}}}
\newcommand{\rps}{\ensuremath{r_{\rm P1}}}
\newcommand{\ips}{\ensuremath{i_{\rm P1}}}
\newcommand{\zps}{\ensuremath{z_{\rm P1}}}
\newcommand{\yps}{\ensuremath{y_{\rm P1}}}

\newcommand{\PS}{\protect \hbox {Pan-STARRS1}}

\def\asec{\char'175 }

\newcommand{\kms}{{\rm km~s}^{-1}}

\begin{document}

\newcommand{\aCfA}{1}
\newcommand{\aUMD}{2}
\newcommand{\aUWar}{3}
\newcommand{\aIUCa}{4}
\newcommand{\aIUCp}{5}
\newcommand{\aNOAO}{6}
\newcommand{\aJHU}{7}
\newcommand{\aFINCA}{8}
\newcommand{\aQUB}{9}
\newcommand{\aIfA}{10}
\newcommand{\aDur}{11}
\newcommand{\aPrince}{12}
\newcommand{\aUSNO}{13}

\title{Towards Characterization of the Type~IIP Supernova Progenitor Population: a Statistical Sample of Light Curves from Pan-STARRS1}
\author{
N.~E.~Sanders,\altaffilmark{\aCfA}
A.~M.~Soderberg,\altaffilmark{\aCfA}
S.~Gezari,\altaffilmark{\aUMD} 
M.~Betancourt,\altaffilmark{\aUWar} 
R.~Chornock,\altaffilmark{\aCfA}
E.~Berger,\altaffilmark{\aCfA}
R.~J.~Foley,\altaffilmark{\aIUCa,\aIUCp}
P.~Challis,\altaffilmark{\aCfA}
M.~Drout,\altaffilmark{\aCfA} 
R.~P.~Kirshner,\altaffilmark{\aCfA} 
R.~Lunnan,\altaffilmark{\aCfA} 
G.~H.~Marion,\altaffilmark{\aCfA} 
R.~Margutti,\altaffilmark{\aCfA} 
R.~McKinnon,\altaffilmark{\aCfA} 
D.~Milisavljevic,\altaffilmark{\aCfA} 
G.~Narayan,\altaffilmark{\aNOAO} 
A.~Rest,\altaffilmark{\aJHU} 
E.~Kankare,\altaffilmark{\aFINCA}
S.~Mattila,\altaffilmark{\aFINCA}
S.~J.~Smartt,\altaffilmark{\aQUB}
M.~E.~Huber,\altaffilmark{\aIfA}
W.~S.~Burgett,\altaffilmark{\aIfA}
P.W.~Draper,\altaffilmark{\aDur}
K.~W.~Hodapp,\altaffilmark{\aIfA}
N.~Kaiser,\altaffilmark{\aIfA}
R.~P.~Kudritzki,\altaffilmark{\aIfA} 
E.~A.~Magnier,\altaffilmark{\aIfA}
N.~Metcalfe,\altaffilmark{\aDur}
J.~S.~Morgan,\altaffilmark{\aIfA}
P.~A.~Price,\altaffilmark{\aPrince}
J.~L.~Tonry,\altaffilmark{\aIfA}
R.~J.~Wainscoat,\altaffilmark{\aIfA} 
and
C.~Waters\altaffilmark{\aIfA} 
} 

\altaffiltext{\aCfA}{Harvard-Smithsonian Center for Astrophysics, 60 Garden Street, Cambridge, MA 02138 USA}
\altaffiltext{\aUMD}{Department of Astronomy, University of Maryland, College Park, MD 20742-2421, USA}
\altaffiltext{\aUWar}{Department of Statistics, University of Warwick, Coventry, UK}
\altaffiltext{\aIUCa}{Astronomy Department, University of Illinois at Urbana-Champaign, 1002 West Green Street, Urbana, IL 61801 USA}
\altaffiltext{\aIUCp}{Department of Physics, University of Illinois Urbana-Champaign, 1110 W. Green Street, Urbana, IL 61801 USA}
\altaffiltext{\aNOAO}{National Optical Astronomy Observatory, 950 North Cherry Ave. Tucson, AZ 85719 USA}
\altaffiltext{\aJHU}{Department of Physics and Astronomy, Johns Hopkins University, 3400 North Charles Street, Baltimore, MD 21218, USA}
\altaffiltext{\aFINCA}{Finnish Centre for Astronomy with ESO (FINCA), University of Turku, V\"ais\"al\"antie 20, 21500 Piikki\"o, Finland}
\altaffiltext{\aQUB}{Astrophysics Research Centre, School of Maths and Physics,Queen’s University, BT7 1NN, Belfast, UK}
\altaffiltext{\aIfA}{Institute for Astronomy, University of Hawaii, 2680 Woodlawn Drive, Honolulu HI 96822}
\altaffiltext{\aDur}{Department of Physics, Durham University, South Road, Durham DH1 3LE, UK}
\altaffiltext{\aPrince}{Department of Astrophysical Sciences, Princeton University, Princeton, NJ 08544, USA}

\email{nsanders@cfa.harvard.edu}

\begin{abstract}
In recent years, wide-field sky surveys providing deep multi-band imaging have presented a new path for indirectly characterizing the progenitor populations of core-collapse supernovae (SN): systematic light curve studies.  We assemble a set of \NIIPtotalF\ $grizy$-band Type~IIP SN light~curves from {\PS}, obtained over a constant survey program of 4~years and classified using both spectroscopy and machine learning-based photometric techniques.  We develop and apply a new Bayesian model for the full multi-band evolution of each light curve in the sample.  
We find no evidence of a sub-population of fast-declining explosions (historically referred to as ``Type~IIL'' SNe).  However, we identify a highly significant relation between the plateau phase decay rate and peak luminosity among our SNe~IIP.  These results argue in favor of a single parameter, likely determined by initial stellar mass, predominantly controlling the explosions of red supergiants.  This relation could also be applied for supernova cosmology, offering a standardizable candle good to an intrinsic scatter of $\lesssim0.2$~mag.
We compare each light curve to physical models from hydrodynamic simulations to estimate progenitor initial masses and other properties of the \PS\ Type~IIP SN sample. We show that correction of systematic discrepancies between modeled and observed SN~IIP light curve properties and an expanded grid of progenitor properties, are needed to enable robust progenitor inferences from multi-band light curve samples of this kind.
This work will serve as a pathfinder for photometric studies of core-collapse SNe to be conducted through future wide field transient searches.
\smallskip
\end{abstract}


\section{INTRODUCTION}
\label{sec:intro} 

Core-collapse supernovae (SNe) mark the explosive deaths of massive stars.  Several independent lines of evidence including explosion modeling \citep{Nadyozhin03,Maguire12,Jerkstrand13,Takats13}, progenitor star photometry
\citep{Li07,Smartt09MNRAS,Walmswell12}, rate statistics \citep{Smith11}, and theory \citep{Heger03,Ekstrom12} combine to suggest a lower main sequence initial mass ($M_{in}$) limit for achieving core collapse of $M_{in}\gtrsim8-12~\rm{M}_\odot$.  Red supergiant progenitor stars in this mass range are known to produce Type~IIP (hydrogen rich) SN explosions, the most common form of core-collapse SN.  The upper mass limit for SNe~IIP progenitors is more uncertain, with stars of $M_{in}\gtrsim16-30~\rm{M}_\odot$ realizing significant mass loss depending on their mass, metallicity, rotation rate, binarity, and other properties; and even more massive stars ending their lives through more exotic explosion mechanisms.  These mass limits for CC-SN progenitor stars have profound implications throughout stellar and galactic astrophysics and cosmology, including as an input to and constraint on models of stellar evolution for massive stars \citep{Groh13,Meynet13}, chemical evolution \citep{Timmes95,Nomoto06,Nomoto13}, supernova feedback in the interstellar medium and galaxy formation \citep{Leitherer92,Stilp13}, and astrobiological planetary sterilization rates \citep{Clark77,Lineweaver04}.

The electromagnetic signatures of these core-collapse explosions are diverse, depending sensitively on the properties of both the core and the outer envelope of the progenitor star at the time of explosion.  Supernovae with hydrogen features detected in their optical spectra are referred to as Type~II SNe, with a variety of subtypes defined by more specific spectroscopic and/or photometric criteria \citep[see e.g.][]{Filippenko97,Li11}.  The most common subclass, Type~IIP, are typified by broad ($\sim10,000~\kms$) hydrogen Balmer P-Cygni spectroscopic features, fast rise times of a few days and optical light curves dominated by a long lived, $\sim100$~day ``plateau'' phase of roughly constant luminosity.  The plateau phase is understood to arise from hydrogen recombination in the ejecta, with cooling temperature balancing the expansion of the blastwave to essentially equilibrate the $R$-band luminosity \citep[see e.g.][]{Kasen09}.  The Type~IIL sub-class is historically designated based on spectroscopic properties similar to SNe~IIP, but faster, ``linearly'' declining optical light curves rather than a long lived plateau.  Type IIb supernovae are classified spectroscopically based on the disappearance of H features and the prominence of He absorptions.  Type~IIb light curves feature slow rise times and rapid decline rates (in each case, a few weeks) typical of Type~I (H deficient) SNe.  The most extreme subclass, Type~IIn, are identified by intermediate width ($\sim10^3~\kms$) H emission features reflecting interaction of supernova ejecta with circumstellar material, and contributions from this interaction can power these explosions to reach extreme luminosities at peak.

The optical evolution of Type~IIP SNe has been explored in light curve studies by a number of authors, including \cite{Patat94,Chieffi03,Hamuy03,Nadyozhin03,Bersten09,Li11,Arcavi12,Anderson14phot,Faran14}.  The relationship between these observables and the properties of SN progenitor stars has been explored in theoretical parameter studies by \cite{Arnett80,Litvinova85,Young04,Kasen09,Dessart13}, and others.  Combining a uniform analysis of a statistical population of Type~IIP supernova light curves with consistent physical models for inferring the properties of their stellar progenitors represents a path forward for characterizing the progenitor population.

Here we describe an analysis of a statistical sample of SN~IIP light curve properties, performed using observations from the Panoramic Survey Telescope \& Rapid Response System 1 survey (\PS, abbreviated PS1).  This represents the first such population analysis of SN~IIP light curves based on a homogeneously-collected and multi-band photometric sample from a wide field optical survey.  In Section~\ref{sec:obs} we describe the PS1 optical observations and follow-up optical spectroscopy program used to construct the light curve sample.  We have developed a novel Bayesian methodology for self-consistently modeling the full population of light curves in the sample and obtaining robust measurements of physically-meaningful light curve parameters (Section~\ref{sec:model}).  We discuss the population wide distributions of these parameters and compare them to previous observational studies (Section~\ref{sec:res}).  By comparison to theoretical light curve models, we recover estimates of the progenitor properties of the objects in our sample, and discuss the limitations of the available models in Section~\ref{sec:prog}. We summarize and conclude in Section~\ref{sec:conc}.  In a companion paper, \cite{Sanders14Unsup}, we apply the PS1 SN~II data presented here as a test case for a hierarchical Bayesian light curve fitting methodology which enables simultaneous modeling of full populations of transient light curves.

\section{OBSERVATIONS}
\label{sec:obs}

\subsection{Pan-STARRS1 imaging}
\label{sec:obs:PS1}

We select a Type~II SN light curve sample from the transients discovered and monitored by PS1 since the initiation of the survey in 2010, consisting of \NphotTot\ relevant photometric data points, \NphotTotDet\ of which are robust transient detections.  PS1 is a high-etendue wide-field imaging system, designed for dedicated survey observations and located on the peak of Haleakala on the island of Maui in the Hawaiian island chain. Routine observations are conducted remotely, from the University of Hawaii--Institute for Astronomy Advanced Technology Research Center (ATRC) in Pukalani.  A summary of details of PS1 operations relevant to SN studies is given in \cite{Rest13PS1Ia}, and we discuss its key features here. 

A complete description of the PS1 system, both hardware and software, is provided by \cite{PS1}. The 1.8~m diameter primary mirror, $3.3^\circ$ field of view, and other PS1 optical design elements are described in \cite{PS1opt}; the array of sixty $4800\times4800$, 0.258\asec pixel detectors, and other attributes of the PS1 imager is described in \cite{PS1cam}; and the survey design and execution strategy are described in \cite{PS_MDRM}.  The PS1 Medium Deep Survey (MDS) consists of 10 pencil beam fields observed with a typical cadence of 3~days in each filter, to a $5\sigma$ depth of $\sim23.3$~mag in $griz$ filters, and $\sim21.7$~mag in the $y$-filter (with observations taken near full moon).

\begin{figure}
\plotone{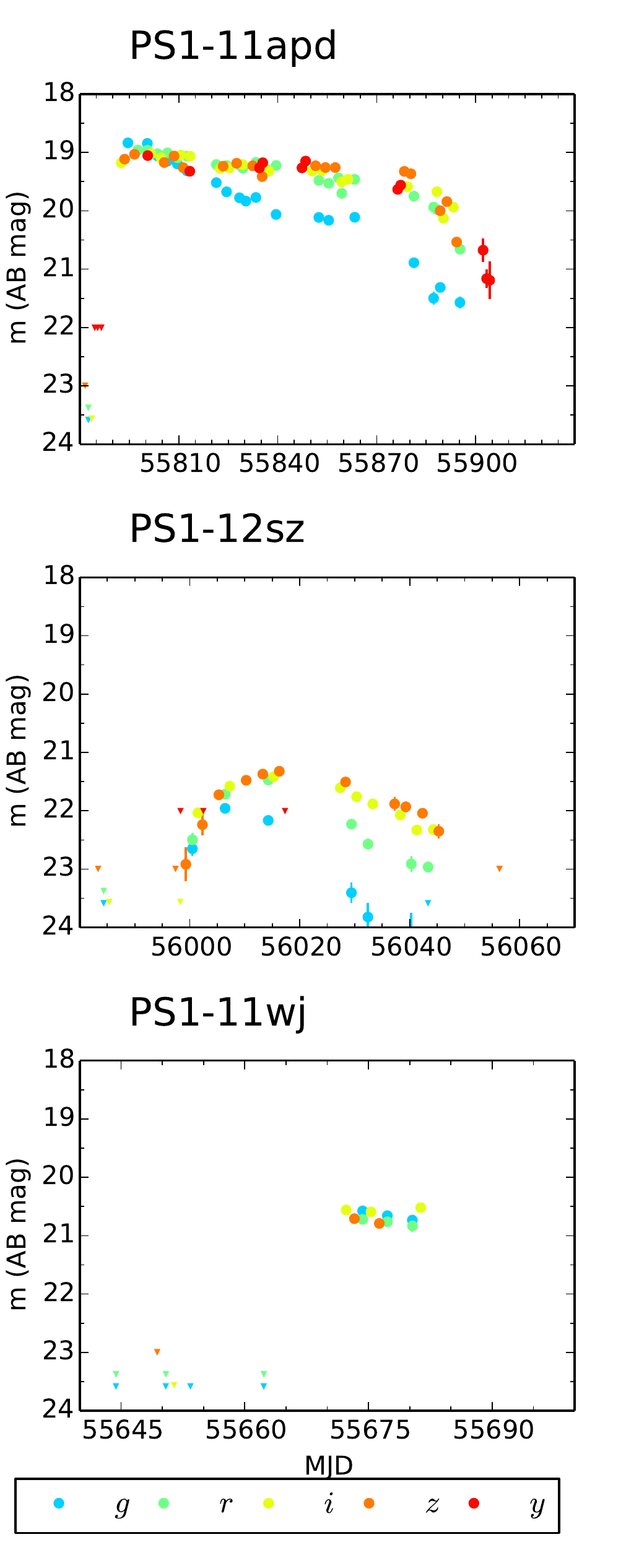}
\caption{\label{fig:MLCsamp}Sample PS1 multi-band light curves from the Type~II SN dataset.  Top: {\PSOthreezerozerotwotwoone}, a bright Type~IIP SN with a light curve sampled from peak through the plateau and the transition to the radioactive decay phase, covering $\sim100$~days.  Middle: {\PSOthreesevenzerothreethreezero}, a Type~IIb supernova which peaked at $\sim21$st magnitude with a well sampled light curve.  Bottom: {\PSOonefivezerosixninetwo}, a Type~IIP SN with photometry sampled only during the plateau phase.  In each plot, the filters are specified by the colors shown in the legend and the triangles represent non-detection upper limits (with values set from the $2~\sigma$ level of the distribution of detected magnitudes from the full photometric dataset in each filter).}
\end{figure}

The PS1 observations are obtained through a set of five broadband filters, which we refer to interchangeably as as \gps, \rps, \ips, \zps, and \yps\ or simply $grizy$ \citep{PS1cal}.  Although the filter system for PS1 has much in common with that used in previous surveys, such as the Sloan Digital Sky Survey \citep[SDSS:][]{York00,SDSS8}, there are important differences. The \gps\ filter extends $200~{\rm \AA}$ redward of $g_{\rm SDSS}$, and the \zps\ filter is cut off at $9200~{\rm \AA}$. SDSS has no corresponding \yps\ filter. Further information on the passband shapes is described in \cite{PS1cal}. Photometry is in the ``natural'' PS1 system, $m = −2.5 \log(\rm flux) + m^\prime$, with a single zero-point adjustment $m^\prime$ made in each band to conform to the AB magnitude scale \citep{Tonry12}.\footnote{The magnitudes quoted throughout this paper are in the AB system, except where explicitly noted.}  We assume a systematic uncertainty of 1\% for our PS1 observations due to the asymmetric PS1 point spread function and uncertainty in the photometric zero-point calibration \citep{Tonry12}.  See Figure~\ref{fig:MLCsamp} for an illustration of the PS1 photometric sampling and the range in data quality.

The standard reduction, astrometric solution, and stacking of the nightly images is done by the Pan-STARRS1 IPP system \citep{PS1_IPP,PS1_astrometry}. The nightly MDS stacks are transferred to the Harvard Faculty of Arts and Sciences ``Odyssey'' Research Computing cluster, where they are processed through a frame subtraction analysis using the \textit{photpipe} image differencing pipeline developed for the SuperMACHO and ESSENCE surveys \citep{Rest05,Miknaitis07,Rest13PS1Ia}.  

\subsection{Optical spectroscopy}
\label{sec:obs:spec}

We begin with a selection of PS1-discovered SNe which were classified as Type~II through our spectroscopic follow-up campaign: \NIItotal\ objects in total.   One object in our sample, {\PSOzeroonezeroonesixthree} (SN~2010aq), has previously been reported on in \cite{Gezari10}.  We note that the PS1 spectroscopic follow-up is not complete; brighter objects, those with longer plateau durations, and those with the highest ratio of SN to underlying galaxy light are most likely to be over-represented in our sample, due to their availability for spectroscopy.

Spectra were obtained using the Blue Channel and Hectospec spectrographs of the 6.5~m MMT telescope (\citealt{BlueChannel,Hectospec}), the Low Dispersion Survey Spectrograph (LDSS3) and Inamori-Magellan Areal Camera and Spectrograph (IMACS; \citealt{IMACS}) of the 6.5~m Magellan telescopes, the Gemini Multi-Object Spectrograph of the 8~m Gemini telescopes (GMOS; \citealt{GMOS}), and the and the Andalucia Faint Object Spectrograph and Camera (ALFOSC) at the 2.6 m Nordic Optical Telescope.  Objects were classified as Type~II by identification of H$\alpha$ emission not associated with the host galaxy.  This selection could possibly include the Type~IIP, IIL, IIb, and Type~IIn subclasses of the SNe~II; we will discuss sub-classification in Section~\ref{sec:mod:class}.  

\begin{figure}
\plotone{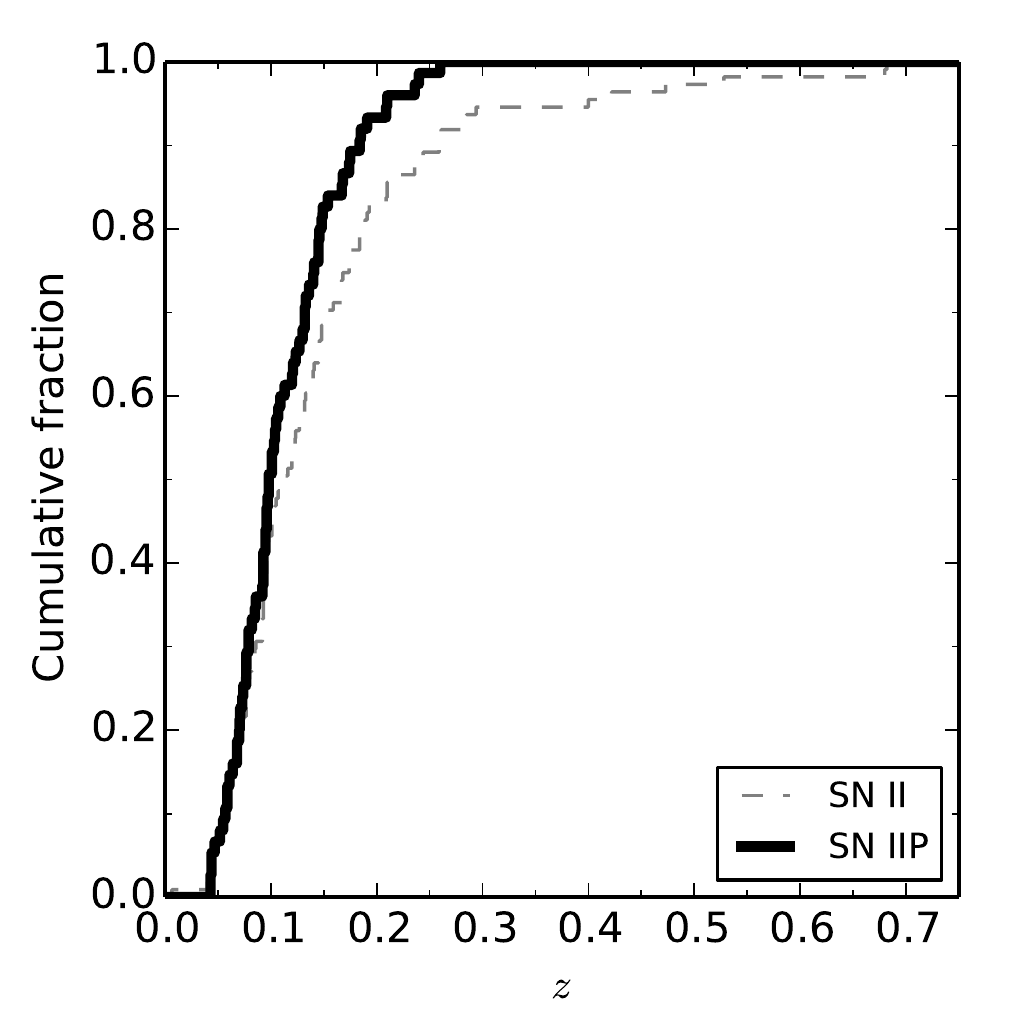}
\caption{\label{fig:zdist}Cumulative redshift distribution for PS1 SNe~II.  The sub-sample classified as SNe~IIP (see Section~\ref{sec:mod:class}) are shown with the thick line.}
\end{figure}

Additionally, we obtain host galaxy redshift measurements for each object from these spectra, which we use for $K$-correction (see Section~\ref{ssec:Kcor}) and to estimate distance.\footnote{We assume a standard $\Lambda$CDM cosmology with $H_0=70$~km~s$^{-1}$ Mpc$^{-1}$, $\Omega_\Lambda = 0.73$, and $\Omega_M=0.27$; \cite{Komatsu11}}  Figure~\ref{fig:zdist} shows the host galaxy redshift distribution of the PS1 SN~II sample, which has [16,50,84]th percentile values of $[\zdistIIAd,\zdistIIAm,\zdistIIAu]$.  The sub-sample classified as SNe~IIP (see Section~\ref{sec:mod:class}) have distribution percentile values of [\zdistIIPd,\zdistIIPm,\zdistIIPu].  This difference reflects a population of luminous, distant SNe~IIn excluded from the sub-sample.

Details of our final SN~IIP sample, as described in Section~\ref{sec:mod:class}, are listed in Table~\ref{tab:obspars}.

\label{sec:obs:PS1SNII}

\subsection{$K$-corrections}
\label{ssec:Kcor}

We use $K$-corrections to account for the difference between the observed and rest frame wavelengths of the light collected through the PS1 filters, as a function of the redshift of each object in our sample.  We define this correction, $K=m_R-M_Q-DM$, where $m_R$ is the observed-frame magnitude, $M_Q$ is the emitted-frame absolute magnitude, and $DM$ is the distance modulus \citep{Hogg02}.  We estimate $K$-corrections using the SN~IIP spectral templates provided by P.~Nugent (based on observations of SN~1999em; \citealt{Gilliland99,Baron04})\footnote{Spectral templates obtained from P.~Nugent at \url{http://supernova.lbl.gov/~nugent/nugent_templates.html}}. We interpolate linearly between the spectral templates to obtain $K$-corrections for arbitrary epochs.  We note that we do not warp the spectral templates to match the observed color of the SN, due to the computational cost involved in doing so dynamically within our probabilistic light curve fitting methodology (see Appendix~\ref{ap:stan}).  By testing the effects of warping the spectral templates on the $K$-correction calculation, we estimate that neglecting this effect will introduce an uncertainty of $\sim0.02$~mag in the $K$-correction per $0.1$~mag of $(V-I)$ color offset, which is much smaller than typical uncertainties in our color estimates.

\begin{figure}
\plotone{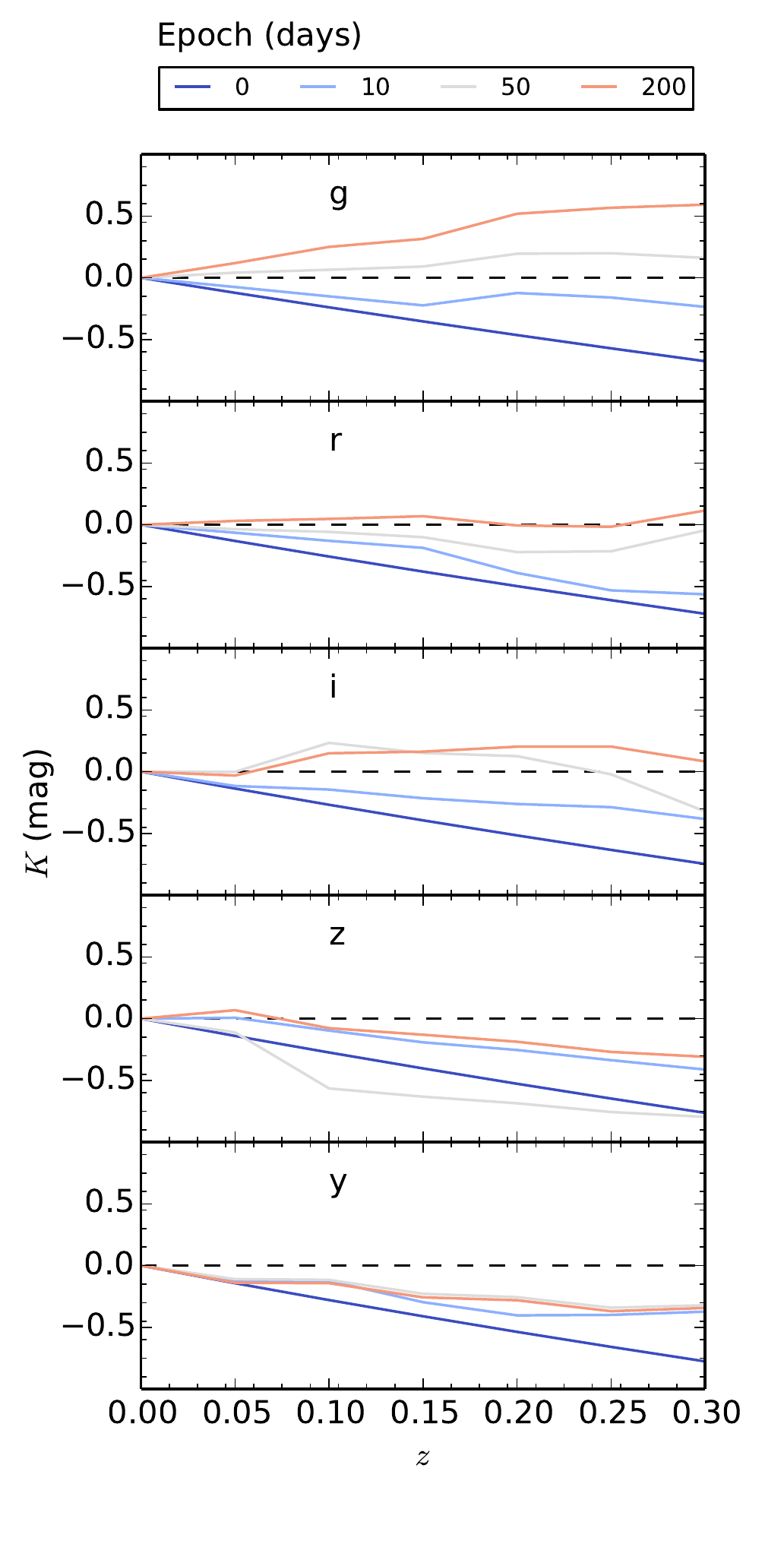}
\caption{\label{fig:Kcor}Estimated SN~IIP $K$-corrections in PS1 filters (different panels), derived as described in the text.  The line colors correspond to different epochs (rest frame days since explosion; key at top).}
\end{figure}

Figure~\ref{fig:Kcor} illustrates the resulting $K$-corrections for an illustrative set of epochs.  The $t=0$ $K$-correction is linear with $z$ because the post-shock breakout cooling phase spectral model is a blackbody; thereafter the $K$-correction evolution becomes more complex, with spectral features evolving and shifting between bandpasses.  The bluer bands have $K$-corrections more highly dependent on time, as the onset of line blanketing in the spectral model depresses the flux blueward of $\sim5000~\rm{\AA}$.

Deviation of the supernova spectral energy distribution from the template spectra will introduce error into our derived $K$-correction.  In particular, we consider here the effect of differences in the SN~IIP continuum shape and line-of-sight extinction, and estimate the magnitude of these effects by simulation on the template spectra.  First, we consider the effect of variation in continuum shape, to first order.  We vary the intrinsic $(V-I)$ color of the supernova spectrum by $\pm0.5$~mag from the stock template, at fixed redshift ($z=0.1$), epoch ($t=50$~days), and with no extinction.  The effect of this variance in spectral shape on the $K$-correction over the full $1$~mag color range induces a variation in the $K$-correction ranging from $0.1$~mag in $g$-band to $0.3$~mag in $y$-band.  Second, we consider the effect of variation in extinction.  We vary the extinction from $A_V=0$ to $2$~mag, at fixed redshift ($z=0.1$), epoch ($t=50$~days), and with the color of the stock template.  The effect of this variance in spectral shape on the $K$-correction over the full $2$~mag extinction range induces a variation in the $K$-correction ranging from $0.1$~mag in $y$-band to $0.2$~mag in $i$-band.  These effects are therefore modest and similar in size, depending on the exact parameters of the photometric observation.

We estimate the typical effect of the joint variation in both continuum shape and extinction by means of Monte Carlo simulation.  We generate $10,000$ random supernova observations with the following conditionally independent distributions of properties meant to approximately mimic the PS1 dataset: redshift drawn from $z\sim\rm{logN}(-2,0.4)$, color deviation drawn from $\Delta (V-I)\sim\rm{N}(0,0.5)$~mag\footnote{Our simulated $(V-I)$ color distribution is meant to provide a conservative accounting of intrinsic spectral deviance between SNe~IIP.  For comparison, \cite{Olivares08} identify an intrinsic $(V-I)$ color dispersion of $<0.1$~mag.}, extinction drawn from $A_V\sim\rm{logN}(0,1)$~mag, and epoch drawn from $t\sim\rm{logN}(4,0.8)$~days, where $\rm{N}$ is the normal and $\rm{logN}$ is the lognormal distribution.  We then consider the distribution of residuals of the derived $K$-corrections from the $K$-correction expected for a zero-extinction and template-standard color object.  In the $r$-band, this distribution has $[5,16,50,84,95]$ percentile values of $[-0.10,-0.01,0.13,0.40,0.75]$~mag.  The distribution is similar in other filters.  This suggests that these effects exert a small ($\sim0.1$~mag) typical bias and introduces a modest additional uncertainty ($\sim0.2$~mag) in the measured supernova photometry.  We therefore neglect these effects in the following work, but note their importance for studies requiring precision photometry.  In the most extreme cases, particularly for the most highly reddened supernovae, the simulations suggest that these effects will cause the brightness to be significantly under-estimated (by $\gtrsim0.7$~mag).  Moreover, the simulated $K$-correction residuals seem to have little correlation between filters, suggesting that the effects do not significantly bias color inference.

\section{LIGHT CURVE MODELING}
\label{sec:model}

In order to consistently compare the photometric properties of the SN~IIPs in our sample, we have produced models of the full, multi-band light curve evolution of each SN using a Bayesian methodology.  We apply weakly informative priors to regularize the shape of the light curve fits to conform to SN~IIP phenomenology.  The product of the modeling is a posterior predictive probability distribution for the luminosity of the SN at every phase, and joint posterior probability distributions for associated light curve parameters.  This methodology enables us to model the full pseudo-bolometric light curve evolution of all objects in our sample from explosion through the radioactive decay phase, regardless of variation in photometric coverage and data quality, while fully accounting for statistical uncertainty.  Throughout, we interpret these marginal parameter probability distributions in the context of the influence of the weakly informative priors to avoid introducing bias on our inference.

\subsection{Parameterized light curve model}
\label{ssec:mod}

To model the light curve, we use a simple, physically-motivated parameterization that captures the essential components of the rise, plateau, and decline phases of the Type~IIP light curve.  Our model is constructed so that the most salient features of the SN~IIP light curve (the plateau duration, peak magnitude, and the luminosity evolution of the radioactive decay phase) are represented by directly interpretable parameters.  

\begin{figure}
\plotone{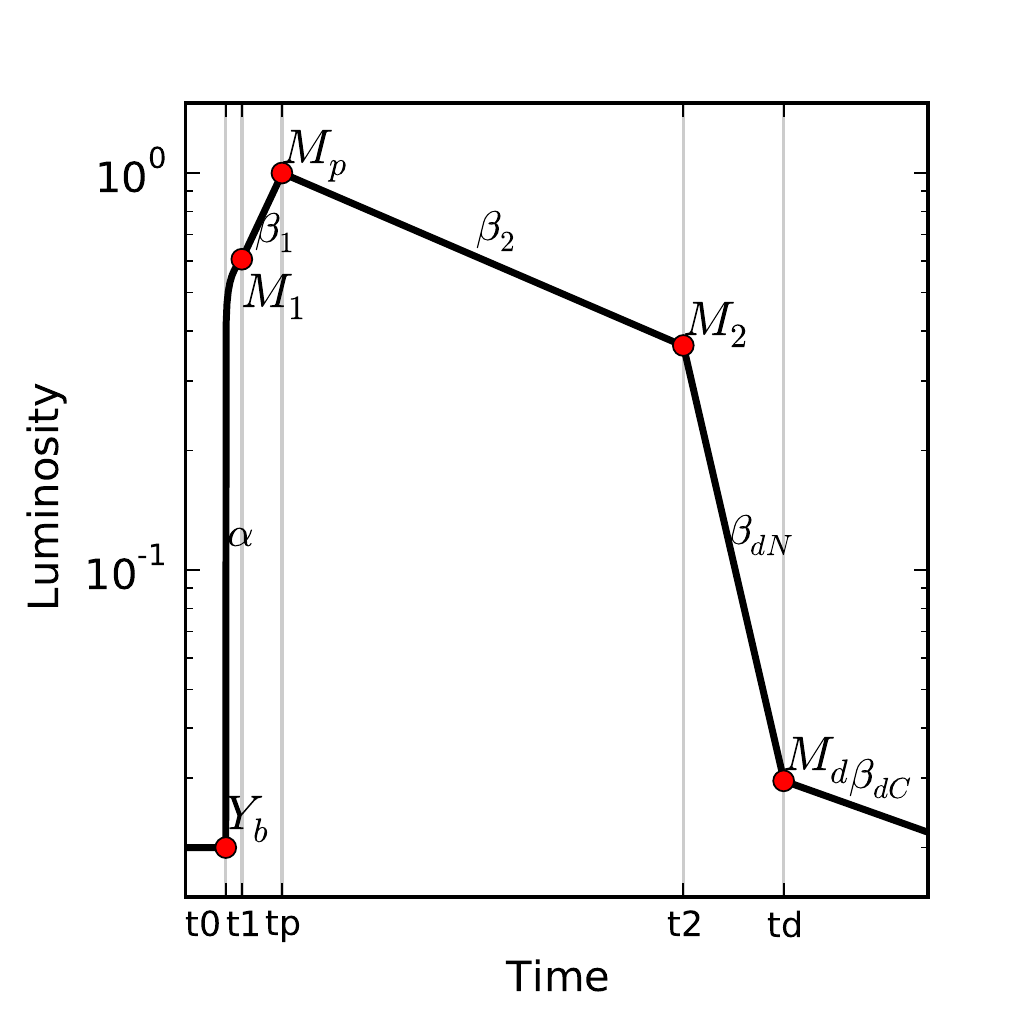}
\caption{\label{fig:modschem}Schematic illustration of the 5-component SN~II light curve model defined in Equation~\ref{eq:model}.  The gray vertical lines denote the duration ($t_x$) between epochs of transition between the piecewise components of the model.  The background level ($Y_b$) and turnover fluxes ($M_x$) are marked and labeled (red points).  The power law ($\alpha$) and exponential ($\beta_x$) rate constant for each phase are labeled adjacent to each light curve segment.}
\end{figure}

The light curve model for the luminosity measured from our PS1 forced photometry, $l$, in a given optical filter, $F$, consists of 5 piecewise components (see Figure~\ref{fig:modschem}): a power law fast rise phase, a plateau phase divided into exponential rising and declining components, a transitional phase representing the end of the plateau, and a post-plateau exponential decay phase:

\begin{align}\label{eq:model}
l[t,\ldots]=\begin{cases}
    0                                      ,\\&\hspace{-14em} \text{if } t<t_0 \\
    M_1 ~ (t/t_1)^{\alpha}          ,\\&\hspace{-14em} \text{if } t_0<t<t_0+t_1 \\
    M_1 \exp(\beta_1 (t-t_1))     ,\\&\hspace{-14em} \text{if } t_0+t_1<t<t_0+t_1+t_p \\
    M_p \exp(-\beta_2 (t-(t_p+t_1)))    ,\\&\hspace{-14em} \text{if } t_0+t_1+t_p<t<t_0+t_1+t_p+t_2 \\
    M_2 \exp(-\beta_{dN} (t-(t_2+t_p+t_1))) ,\\&\hspace{-14em} \text{if } t_0+t_1+t_p+t_2<t\\&\hspace{-10em}<t_0+t_1+t_p+t_2+t_d \\
    M_d \exp(-\beta_{dC} (t-\\\hspace{4em}(t_d+t_2+t_p+t_1))) ,\\&\hspace{-14em} \text{if } t_0+t_1+t_p+t_2+t_d<t \\
\end{cases}  \notag \\
\end{align}

These parameters have the following definitions and interpretations.  The time parameters are defined where $t$ is the Modified Julian Date (MJD) epoch of an observation, $t_0$ is the epoch of explosion, $t_1$ is the rest frame duration of the power law rise phase, $t_p$ is the duration of the exponential rise phase (ending at peak flux), $t_2$ is the duration of the falling component of the plateau phase, and $t_d$ is the duration of the transitional phase.  The flux parameters are defined such that $M_1$ is the flux at the transition from the power law to the exponential rise phases, $M_p$ is the peak flux, $M_2$ is the flux at the end of the plateau phase, and $M_d$ is the flux at the transition to the Co decay-dominated phase.  The rate parameters are defined such that $\alpha$ is the power law rise slope, $\beta_1$ is the exponential rate constant during the rising phase of the plateau, $\beta_2$ is the rate constant during the declining phase of the plateau, $\beta_{dN}$ is the exponential decline rate of the transition phase following the plateau, and $\beta_{dC}$ is the exponential decay constant corresponding to $^{56}$Co to $^{56}$Fe decay.  Each parameter is defined independently for each photometric filter, with the exception of $t_0$.  For numerical convenience, we define $l$ in arbitrary scaled units relative to the absolute magnitude $M$, such that $M = -2.5 \log_{10}(10^7 \times l)$.

Note that, in order for the light curve model to be continuous, not all of the parameters may be independent.  In particular, for each filter,

\begin{align}
M_1 &= M_p / \exp(\beta_1 t_p) \\
M_2 &= M_p / \exp(-\beta_2 t_2) \\
M_d &= M_2 / \exp(-\beta_{dN} t_d)
\end{align}

Furthermore, we note that the post-peak decay rate $\beta_2$ is directly related to the quantity $\Delta m_{15}$, the decline in magnitudes of the light curve in the 15~days following peak:
    
\begin{align}
\Delta m_{15} &= \frac{15\times2.5}{\log_e{10}}~ \beta_2  \sim  16.3 ~\beta_2
\end{align}

Our model is similar to the linear segmented light curve fitting approach of e.g. \cite{Patat93}, but (as we will discuss in Section~\ref{sec:mod:fit}) our fitting methodology for the knot properties is fully probabilistic rather than manual.  We prefer this piecewise analytic formulation to the additive components model used by e.g. \cite{Olivares08}.  The piecewise parameters will in principal have weaker covariance, therefore reducing the posterior curvature and increasing the efficiency of Markov Chain Monte Carlo (MCMC) methods for sampling from the posterior.  While this parameterization is designed to capture the phenomenology of Type~IIP SNe, it is in practice quite flexible, as we obtain reasonably descriptive fits to the light curves of other Type~II SNe (e.g. SNe~IIn and IIb).

\subsection{Fitting methodology}
\label{sec:mod:fit}

We estimate the posterior distributions of these model parameters using a MCMC method.  We employ the C++ library \textit{Stan} \citep{STAN}, which implements the adaptive Hamiltonian Monte Carlo (HMC) \textit{No-U-Turn Sampler} of \cite{NUTS}.  For each multi-band SN light curve, we use \textit{Stan} to return {1000}~samples ({250} samples each from {4} independent MCMC chains) from the posterior distribution of the model.\footnote{The full \textit{Stan} code for our statistical model is discussed in Appendix~\ref{ap:stan}.}

In addition to the light curve parameterization outlined in Section~\ref{sec:model}, our \textit{Stan} model includes certain features representing the data acquisition process.  To account for uncertainty in the PS1 background template subtractions, we fit for the background level in each filter using an independent set of luminosity parameters, $Y_b[F]$, and an intrinsic model variance, $V[F]$.  We pre-compute $K$-correction curves for the redshift of each object in our sample (see Section~\ref{ssec:Kcor}), and apply them to the model during the likelihood calculation using the phases corresponding to the sampled explosion date at each step in the MCMC chain.

\begin{figure*}
\plotone{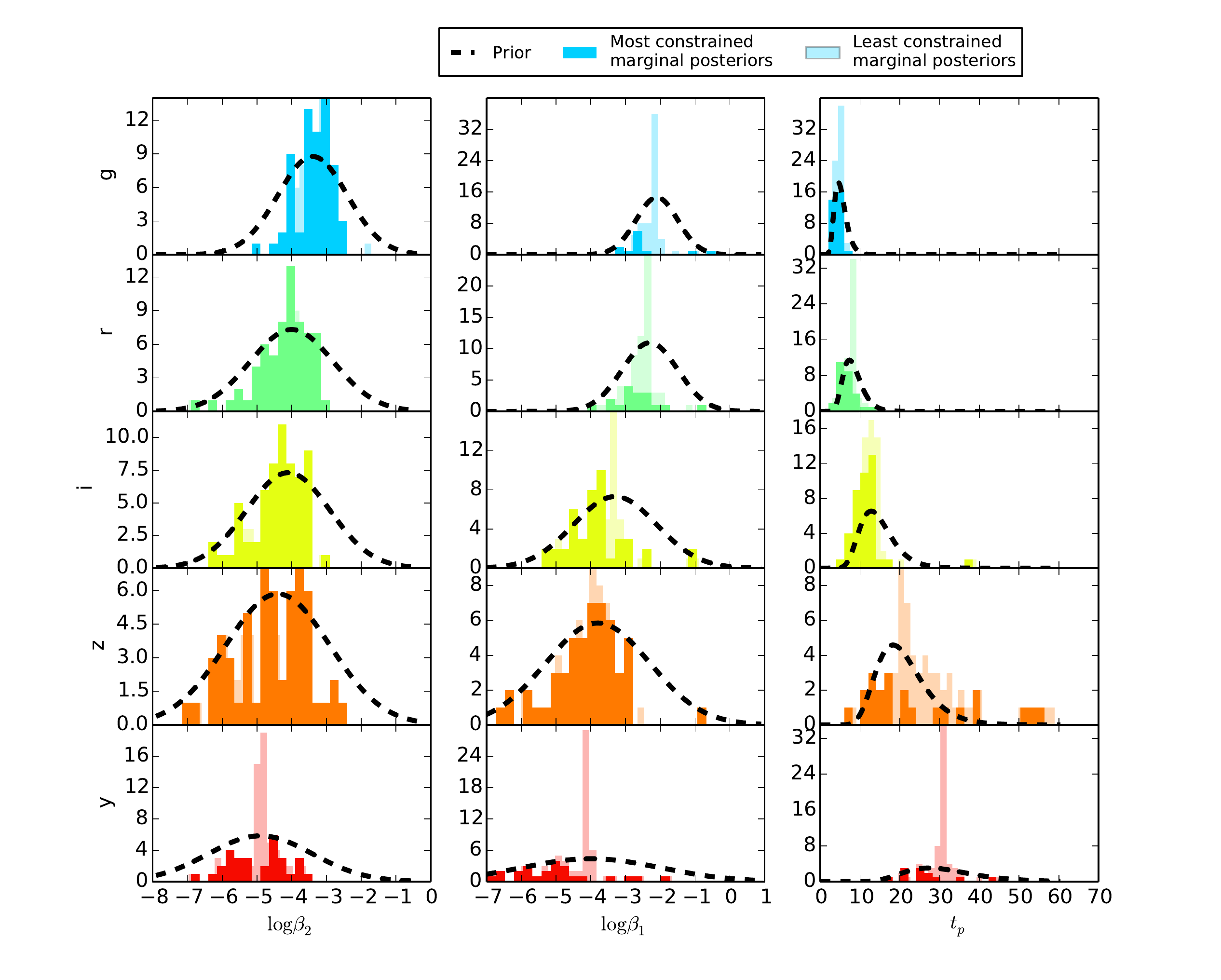}
\caption{\label{fig:pfprior}Illustration of prior distribution construction for parameters of the Bayesian SN~IIP light curve model.  Filters are displayed by row; parameters by column.  Only parameters with priors defined per-filter are shown.  The bars show the distribution of the marginalized posterior medians for the fitted light curve model parameters.  The least constrained posterior medians (with variance 80\% or more of the variance in the prior) are shown with the faded bars, and more constrained posterior medians are shown with darker bars.  The chosen prior distribution is shown with the dashed lines.  Posteriors fitted exclusively with information from the prior (no constraint provided by the data; e.g. the least constrained marginal posteriors) would appear exactly at the position of the prior mean.}

\end{figure*}

\begin{figure}
\plotone{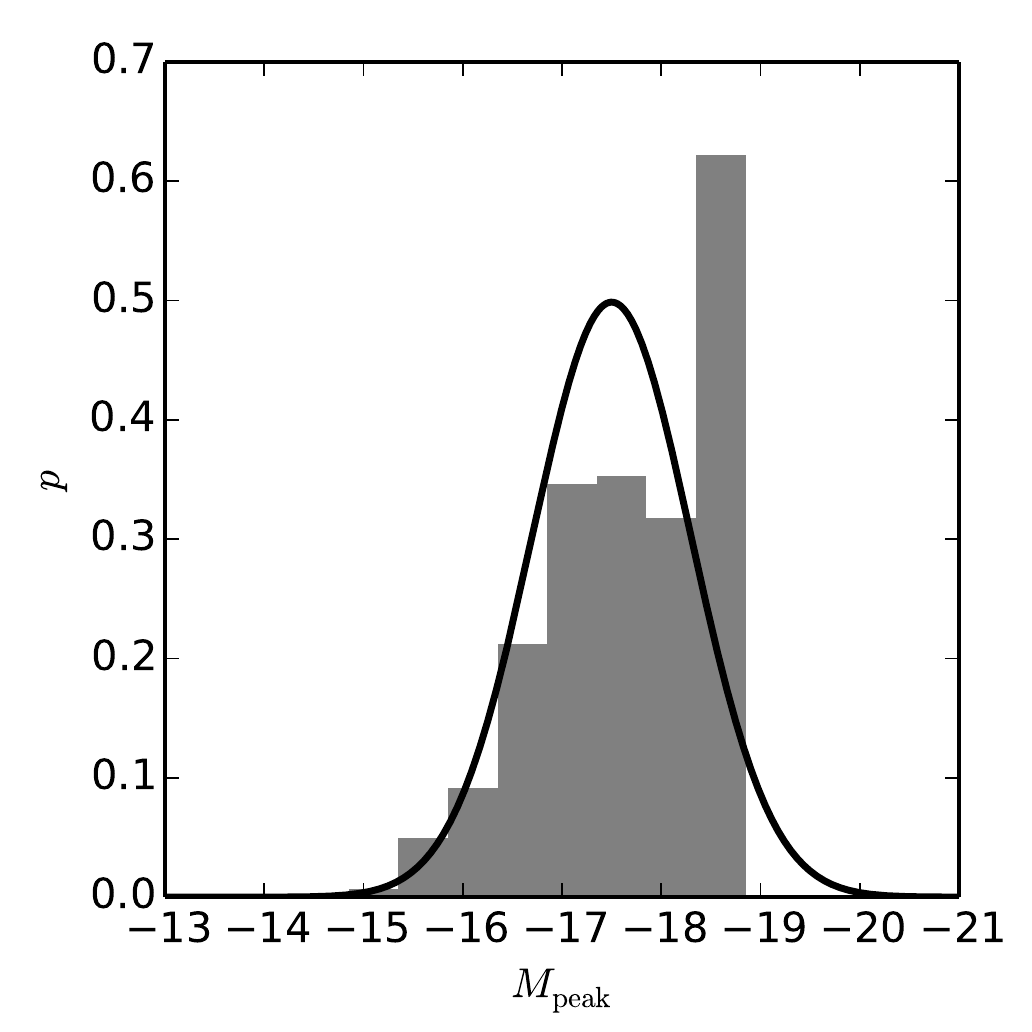}
\caption{\label{fig:mpprior}The prior adopted for the SN~II peak magnitude distribution (solid line) compared to the observed SN~II luminosity function determined by \cite{Li11} for an ideal, magnitude-limited survey with a 1~day cadence (their Figure~10) is shown with gray bars.}
\end{figure}

We employ weakly informative priors (see e.g. \citealt{Gelman08}) to regularize the fitted models to the characteristic SN~IIP light curve shape.  These prior distributions are constructed to more than encompass the observed variance in parameter values from well-identified light curves in the dataset.  The \textit{Stan} model in Appendix~\ref{ap:stan} fully defines the fixed prior distributions we employ, and we discuss their most salient properties here.  Figure~\ref{fig:pfprior} illustrates the procedure for constructing the prior distributions which we have defined individually per filter in the model.  For $t_2$, we apply a (lognormal) $\log\rm{N}(log(100), 0.3)$~day prior.  Together with our prior on $t_p$ (Figure~\ref{fig:pfprior}), this effectively provides a prior on the plateau duration which has $[5,50,95]$th percentile values of $[70,110,180]$~days in $i$-band (and similar in other bands).  This prior is constructed to be weakly informative, having significantly more variance than previously reported SN~IIP plateau duration distributions \citep{Arcavi12} and covering a range of values similar to that predicted from theoretical modeling \citep{Kasen09}.  Perhaps the most informative prior we apply is that on the peak-magnitude distribution, which is observationally motivated.  A strong prior is desirable for this parameter because it has the effect of regularizing the light curve shape in cases where photometric observations before peak are not available.  We set this prior to $N(-17.5~\textrm{mag},0.8~\textrm{mag})$ to approximately match the Type~II SN luminosity function predicted for an $R$-band magnitude-limited survey by \cite{Li11} based on observations from the Lick Observatory Supernova Search (LOSS; Figure~\ref{fig:mpprior}).\footnote{Note that \cite{Li11} treat the SN~II population as a mixture of relatively low luminosity and relatively luminous SNe (referred to as Type~IIP and IIL, respectively), leading to a bi-modal luminosity function as in Figure~\ref{fig:mpprior}.  Here we assume a uni-modal Gaussian prior wide enough to encompass both modes of the \cite{Li11} luminosity function, and investigate the possibility of multi-modality in the observed properties of the SNe~II in Section~\ref{sec:res:IIL}.}

In total, including the independent parameters for each light curve segment and filter, the fitted model for each SN has 61 independent parameters (12 per filter and $t_0$).  We reproduce the fitted light curve parameters for each object in our sample in Table~\ref{tab:lcpars}.

\subsection{Fitting validation}
\label{sec:mod:valid}

Following execution of \textit{Stan} and posterior sampling, we reconstruct a posterior predictive distribution for each light curve from the model parameter samples \citep{BDA3}.  This distribution represents the probability for the physical supernova light curve to have the luminosity $l_m$ on a grid of times $t_m$.

\begin{figure}
\plotone{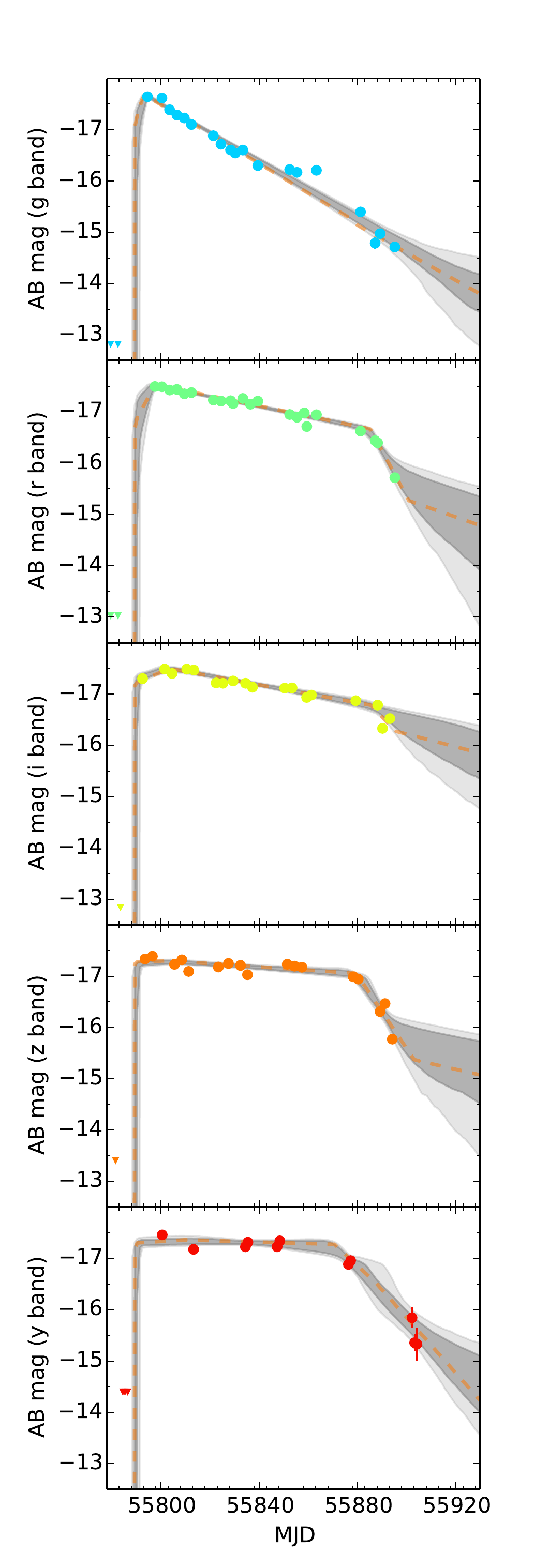}
\caption{\label{fig:modMLC}An example of the 5-component SN~II light curve model fit to the SN {\PSOthreezerozerotwotwoone} in each of the $grizy$ bands (top to bottom).  The PS1 photometry and uncertainties in each of five optical bands are displayed by the circles and errorbars.  The model posterior distribution is displayed by the shaded intervals; the shading boundaries correspond to the $[5,16,84,95]$th percentile values of the posterior.  The maximum likelihood model in each band is displayed by the dashed line.  $K$-corrections have been applied to both the data and model, but not reddening correction.  The triangles denote photometric upper limits.}
\end{figure}

Figure~\ref{fig:modMLC} illustrates a posterior predictive check for the model fit to the object {\PSOthreezerozerotwotwoone}, whose well-sampled observations illustrate both the strengths and weaknesses of the 5-component model.  The rising phase in each band is well-constrained by the tightest pre-explosion limit, in the $y$-band.  The two-component (rising then falling) plateau phase provides an accurate description of the data in the $i$-band, but not in the $g$-band, where the evolution of iron line blanketing (see e.g. \citealt{Kasen09}) causes a behavior not captured by the model (falling then rising/plateauing).  As is characteristic of the PS1 data, the $y$-band light curve is more poorly sampled and higher in variance than the other bands.  In general, the model parameter inferences are therefore more reliable and directly interpretable in the $r$, $i$, and $z$ bands.

\begin{figure}
\centering{
\plotone{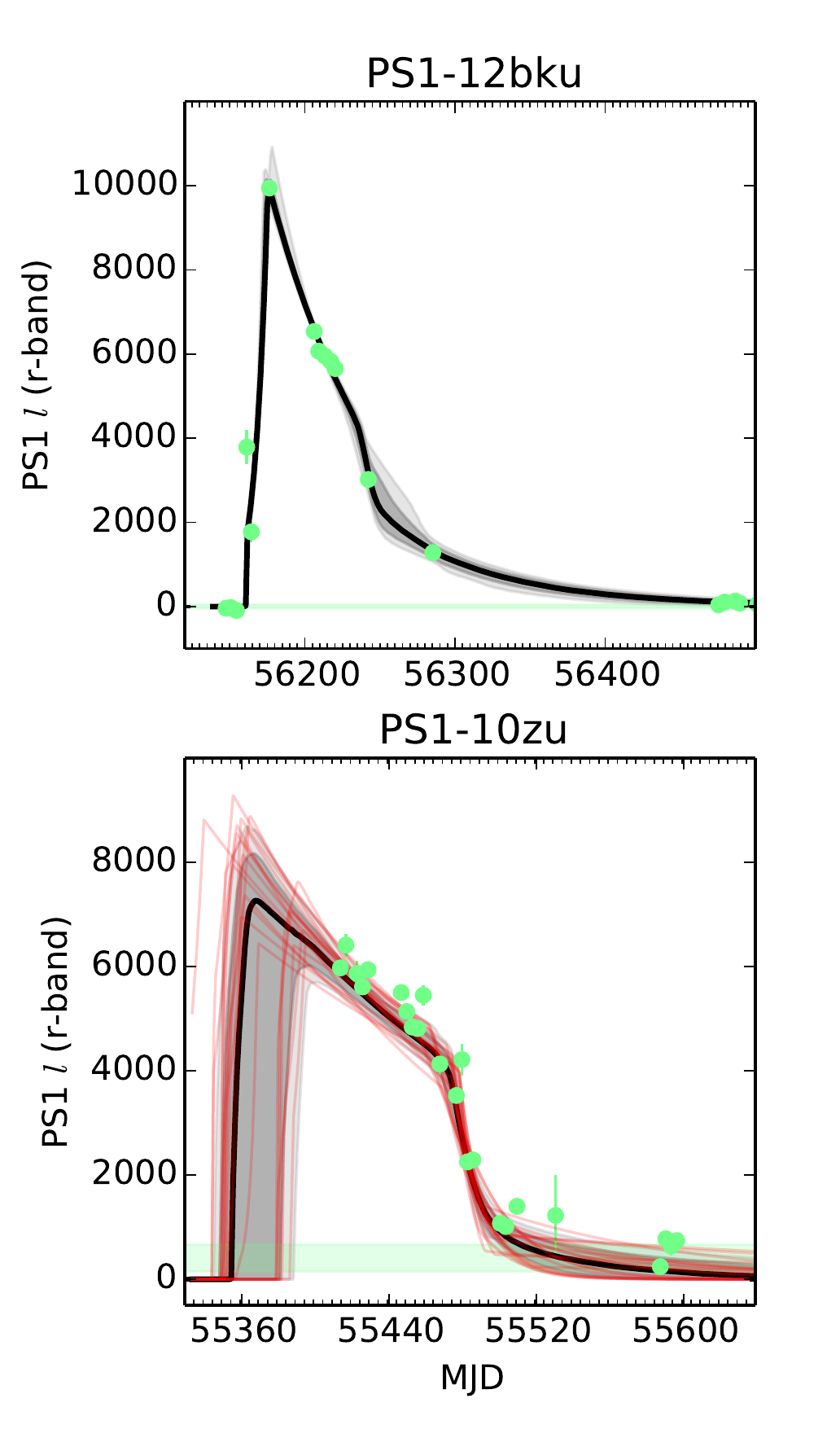}
}
\caption{\label{fig:modex}Examples of fitting the 5-component SN~II light curve model defined in Equation~\ref{eq:model} over a range in light curve quality.  Top: $r$-band photometry for the SN~IIP {\PSOfourtwozerothreeninethree} (green points and errorbars), showing clear detections during each of the five phases on the light curve model.  The shaded areas show the $1$ and $2~\sigma$ confidence intervals for the posterior distribution of the model (see Section~\ref{sec:mod:fit}) and the solid black line shows the median of that confidence interval.  The horizontal green bar shows the range fitted for the zero-point flux offset.  Bottom: The same, for {\PSOzerosixoneoneninesix}.  In this case, no observations constrain the rise phase, and models with a diversity of plateau durations and rise behaviors are explored by the MCMC chain (a sampling of these models is shown in red).}
\end{figure}

Figure~\ref{fig:modex} illustrates fitting behavior for a well-sampled and a poorly-sampled PS1 light curve.  The top panel shows {\PSOfourtwozerothreeninethree}, a SN with observations from the rise phase through the end of the plateau.  Note that the nonlinear behavior of the light curve during the rise, due to shock breakout and post-shock breakout cooling envelope effects (see e.g. \citealt{Nakar10,Rabinak11}) are not captured by the power law rise model, but the explosion epoch itself is fit accurately in the fitted model.  {\PSOzerosixoneoneninesix} is shown in the bottom panel, an object which apparently exploded during the gap between observing seasons.  With observations only available starting a few weeks before the end of the plateau phase, and no direct constraints on the explosion epoch, the model fit includes a range of possible explosion dates reflecting the prior distributions for the plateau duration and peak magnitude. 

\subsection{Type II SN sub-classification}
\label{sec:mod:class}

We apply a supervised machine learning methodology (see e.g. \citealt{Faraway14}) to classify the Type~II SNe in this sample into their appropriate subclasses and, in particular, to identify SNe~IIP for further analysis.  Our approach is to first apply ``expert knowledge'' to manually label a subset of clearly-classifiable objects from our sample of \NIItotal\ SNe~II, then we uniformly measure light curve features using the methodology described in Section~\ref{ssec:mod}, then train a Support Vector Machine (SVM) classifier to a deliberately-selected subset of these features, and finally apply that classifier to assign labels to the remaining objects in the sample.  In general, spectral classification of distant ($z\gtrsim0.1$) transients is complicated by spectral evolution (see e.g. \citealt{Milisavljevic11ei} for a discussion of SNe~IIb), exacerbated by the observational cost of obtaining a well sampled spectral series, and the difficulty in robustly identifying spectral features, exacerbated by the low signal-to-noise of the spectroscopy relative to nearby objects.  The methodology discussed below is intended to capitalize on photometric information to help overcome these limitations.

Our procedure is as follows.  We begin by manually assigning SN~IIP, IIb, IIn, and ``II?'' (undetermined) classifications to each object in the sample by inspection of their light curves and optical spectra.  See e.g. \cite{Li11} for a discussion of the photometric and spectroscopic properties of SN~II sub-classes.  We do not establish a label for SNe~IIL separate from SNe~IIP, but discuss that sub-class later in Section~\ref{sec:res:IIL}.  We assign SN~IIn classifications to \NIIntotal~objects based on the detection of intermediate-width H emission features and/or exceptional luminosity ($\lesssim -20$~mag); SN~IIb classifications to \NIIbtotal~objects based on the detection of He features (in addition to broad H features) in the optical spectrum and/or a SN~I-like light curve, with slow rise and fast decline rate; and SN~IIP to \NIIPtotal~objects based on the presence of a broad $H\alpha$ spectral feature and/or a clearly-exhibited plateau-like light curve.  Finally, we liberally apply the SN~II? classification to \NIIqtotal~objects for which we do not have sufficient data to assign a classification and/or do not neatly fit the preceding criteria.  

We apply our Bayesian light curve model (Section~\ref{ssec:mod}) to uniformly measure features of the light curves of every SN~II in our PS1 sample.  To mirror our expert knowledge, we focus on the peak absolute magnitude ($M$), the rise time to peak magnitude ($t_p$), and the post-peak decay rate ($\beta_2$) as features capable of distinguishing objects in these subclasses.  We identify the $i$ band as the filter in which these features are most discriminating.  For self-consistency, we apply the SN~IIP theoretical $K$-corrections (Section~\ref{ssec:Kcor}) to every object in the sample.

\begin{figure}
\centering{
\plotone{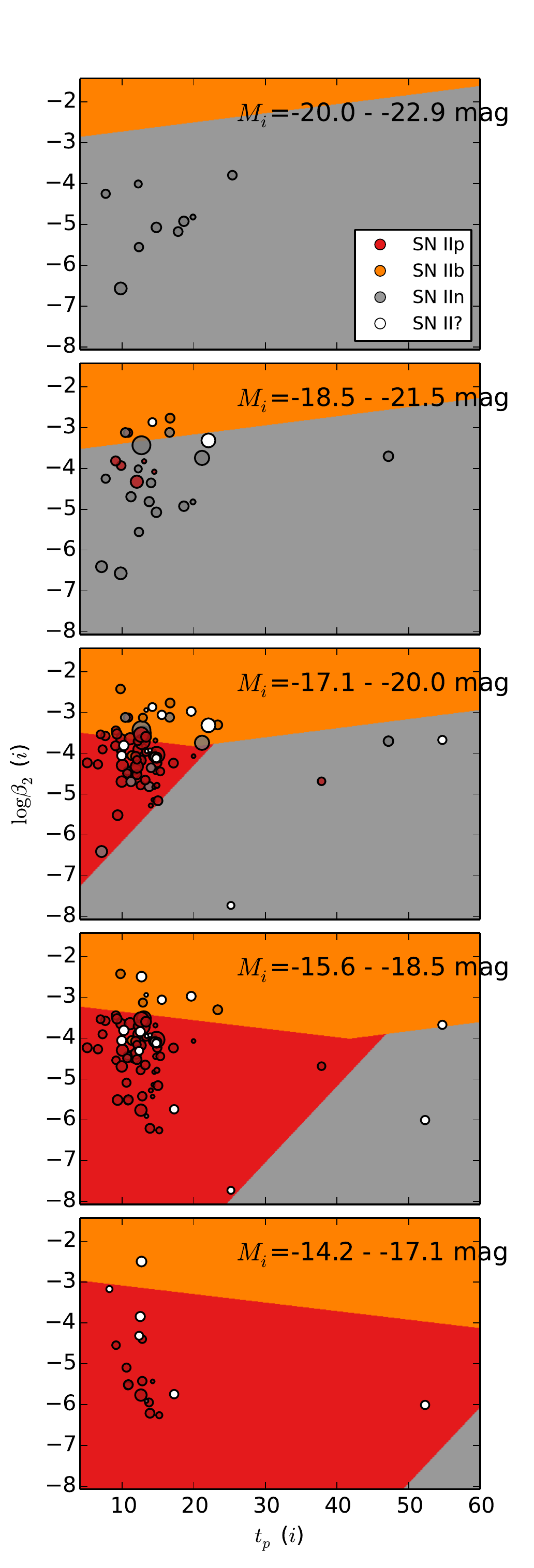}
}
\caption{\label{fig:svm}Support Vector Machine classifier trained to the SN~II light curve feature data.  The subplots display 5 slices of the 3-dimensional feature space ($\beta_2,t_p,M_{i}$), representing different bins of absolute $i$-band magnitude.  The shaded regions indicate the labeling regimes identified by the classifier, and the points represent observed objects from the sample of different manually-assigned sub-classes.  The colored points represent the training set, and the white (``SNII?'') points represent the objects to be classified.  The point size reflects the uncertainty in the feature posterior distribution for each object, with larger points indicating smaller uncertainty.}
\end{figure}

We then fit a linear SVM classifier \citep{scikit-learn} to the features of the training set.  The SVM consists of a series of hyperplanes that optimally divide the feature space to discriminate the training set based on their class labels.  The results of this fit are displayed in Figure~\ref{fig:svm}.  In summary, the classifier identifies SNe~IIn as luminous (primarily $M_i < -18$~mag) and slowly-evolving ($\log \beta_2 \lesssim-3$), SNe~IIb as intermediate-luminosity and quickly evolving, and SNe~IIP as slowly-evolving and spanning a range from sub-luminous ($M_i > -17$~mag) to luminous ($M_i \sim -20$~mag), accurately reflecting our expert knowledge applied in construction of the training set.  Disagreement between object location and shaded labeling in Figure~\ref{fig:svm} is partially due to variance in the observational data, which limits the efficiency of the classifier, and partially due to the coarse binning of the display slices.  We have configured the classifier with the penalty parameter $C=1$ and class weights calculated from their representation in the training set.

Finally, we apply the classifier to the unclassified (SNII?) objects to infer their subclasses.  While ``SN~IIb'' and ``IIn'' classifications can only be assigned by spectroscopy, by definition, here we are using photometric information to supplement available spectroscopy for the purpose of predicting the most likely spectroscopic sub-classification. The classifier identifies \SVMClP\ of these objects as SNe~IIP, which we hereafter incorporate in our SN~IIP sample; \SVMClb\ as SNe~IIb; and \SVMCln\ as SNe~IIn.

When applied to the training set, the classifier achieves \SVMEffAll\% accuracy across all sub-types, and \SVMEffIIp\% accuracy for SNe~IIP.  It mistakenly classifies SNe~IIb and IIn as SNe~IIP only \SVMFPIIp\% of the time (the false identification rate), and mistakenly classifies SNe~IIP as other classes only \SVMFNIIp\% of the time (the false negative rate).  These statistics imply that, of our sample of \NIIqtotal~SNe~II?, $\sim3$ may be incorrectly classified as SNe~IIP and $\sim3$ legitimate SNe~IIP may be inadvertently excluded.  In combination with our sample of \NIIPtotal\ securely-identified SNe~IIP, this level of false identification would imply a $\sim4\%$ contamination rate within the sample.  Given the relatively low S/N of the SN~II? objects, the classifier may perform more poorly than on the training dataset; but for the same reason, the influence of these objects on our light curve parameter inferences will be correspondingly suppressed.

\section{RESULTS}
\label{sec:res}

In this section, we present the results of our analysis of the observational properties of the PS1 SN~IIP light curves.  The parameters calculated for individual objects are summarized in Table~\ref{tab:ppars}.

\subsection{SN~IIP Light Curve Template}
\label{sec:res:stack}

\begin{figure*}
\plotone{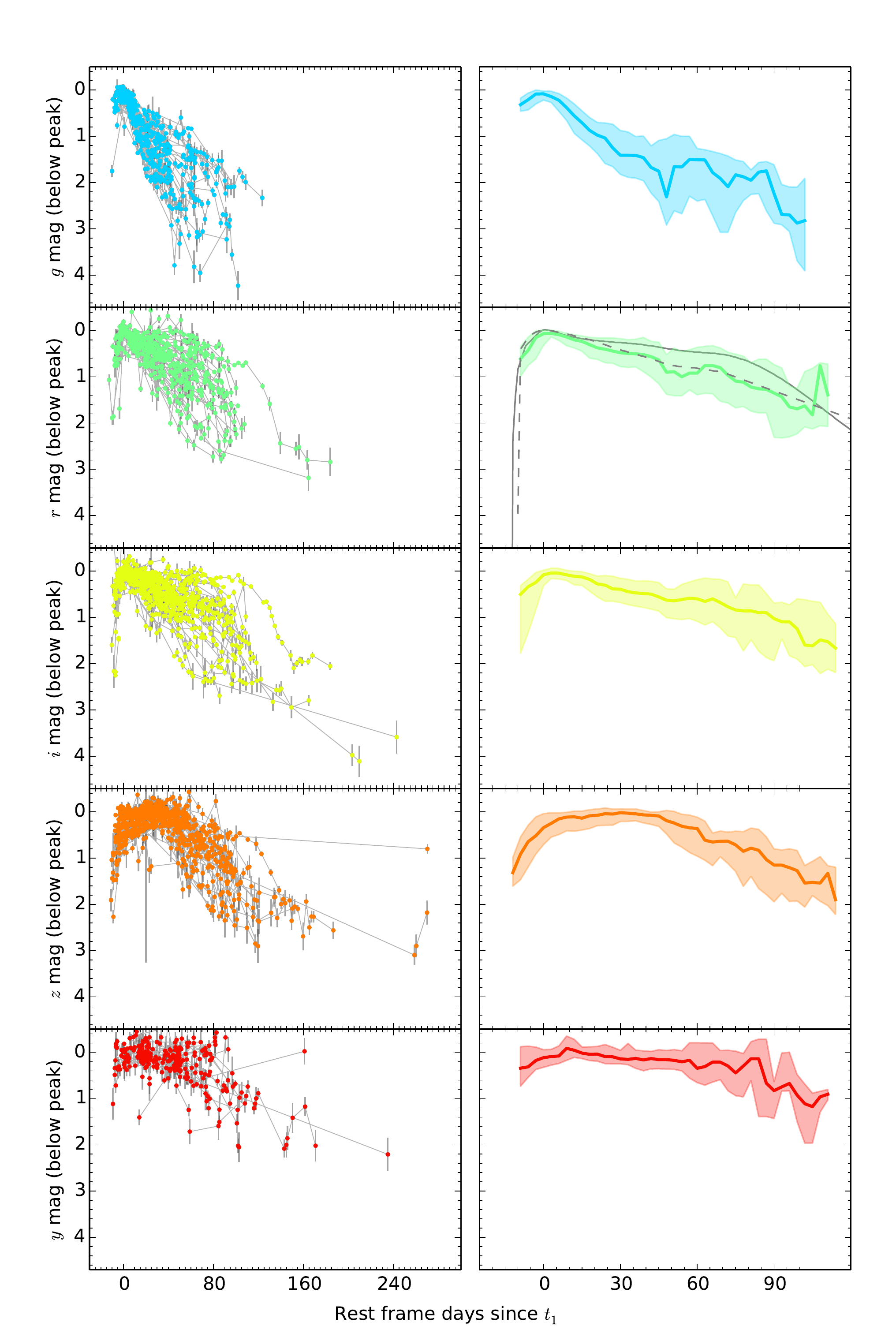}
\caption{\label{fig:stackLC}Left: Stacked SN~IIP light curve constructed from objects in the PS1 sample.  The magnitudes are shown relative to the fitted value of $M_{peak}$ (see Section~\ref{ssec:mod}).  Right: Template light curves derived from the stacks, zoomed in relative to the stacked light curves.  The solid line shows the median value of stacked photometry at each reference epoch, and the shaded region shows the $1\sigma$ (16-84th percentile) range.  The grey lines displayed against the $r$-band curve show the unfiltered SN~IIP (solid line) and IIL (dashed line) templates from \cite{Li11}.  $K$-corrections have been applied using the method described in Section~\ref{ssec:Kcor}, including uncertainties accounting for the range of epochs allowed for each photometric point by the model fit.}
\end{figure*}

We construct $grizy$ SN~IIP template light curves based on scaled stacks of our PS1 photometric sample.  Figure~\ref{fig:stackLC} (left) shows these stacked SN light curves.  In each band, we only include objects with a $1\sigma$ uncertainty in the peak magnitude of $<0.5$~mag and with an uncertainty in the plateau epoch of $<10$~days, as estimated from the marginal posterior of the fitted light curve model.  In total, we include $[\stackNacceptg,\stackNacceptr,\stackNaccepti,\stackNacceptz,\stackNaccepty]$ objects in the $[g,r,i,z,y]$-band templates.

Figure~\ref{fig:stackLC} (right) shows the template light curves constructed from these data.  The template magnitudes are calculated based on the range of observed photometry within a moving window of 6~rest frame days in width, between $\sim-10$--$110$~days from peak magnitude.  Template values are only reported if at least 4 photometric observations fall in the window.  In total, we include $[\stackNtemptotalg,\stackNtemptotalr,\stackNtemptotali,\stackNtemptotalz,\stackNtemptotaly]$~observations in constructing the $[g,r,i,z,y]$-band templates.  The templates are reproduced in Table~\ref{tab:templ}.

\subsection{Search for an SN IIL sub-population}
\label{sec:res:IIL}

Past studies have varied widely in their interpretation of a putative, fast-declining sub-class of SNe~II commonly labeled ``SNe~IIL.''  The SN~IIL classification was coined by \cite{Barbon79}, in recognition of a set of 6 Type~II SNe with an unusually fast decline rate of $\sim0.05~\rm{mag}~\rm{day}^{-1}$ in the $B$-band, constituting $\sim26\%$ of their SN~II sample.  From a theoretical perspective, a fast declining SN~II suggests a less massive H envelope and consequently less sustained power contribution from H~recombination.  Likely explanations for diversity in H envelope mass among SN~II progenitors include variation in progenitor initial mass, with SNe~IIL likely to arise from either progenitors with intrinsically less massive envelopes (e.g. initial masses of $\sim7-10~\rm{M}_\odot$; \citealt{Swartz91}) or from more massive stars which have been stripped by strong radiation driven winds ($\gtrsim18~\rm{M}_\odot$; \citealt{Smith11}), and/or stars which have been partially stripped by interaction with binary companions \citep{Nomoto95}.

Observationally, no widely adopted criteria for SN~IIL classification have emerged.  \cite{Patat94} divided their SN~II light curve set into ``linear'' and ``plateau'' classes based on the  \cite{Barbon79} criterion, but they describe the decline rate distribution of these objects as continuous and find no statistical evidence for two modes in the distribution.  In their population study of the LOSS SNe, \cite{Li11} define SN~IIL as having SN~IIP-like spectroscopic features, but exhibiting an $R$-band decline of $>0.5$~mag after explosion.  By their selected criterion, they identify 7 unambiguous SNe~IIL among 81 SNe~II, and they determine SNe~IIL to represent 10\% of the volume-limited SN~II population.  They find that the SNe~IIL are on average over-luminous compared to SNe~IIP, and therefore would represent $\sim25\%$ of a magnitude-limited SN~II survey sample.  By inspection of a set of 22 SN~II light curves from the CCCP and the literature, \cite{Arcavi12} identify a set of 5 objects which they describe as SNe~IIL.  Their selection corresponds to a decline rate criteria of $\gtrsim0.3$~mag within 50~days of explosion.  They suggest that these objects form a class ``distinct'' from SNe~IIP, and that the two classes together do not form a continuum of decline rates.  Most recently, \cite{Anderson14phot} have characterized the decline rate distribution of hydrogen-rich SNe as continuous.

\begin{figure}
\plotone{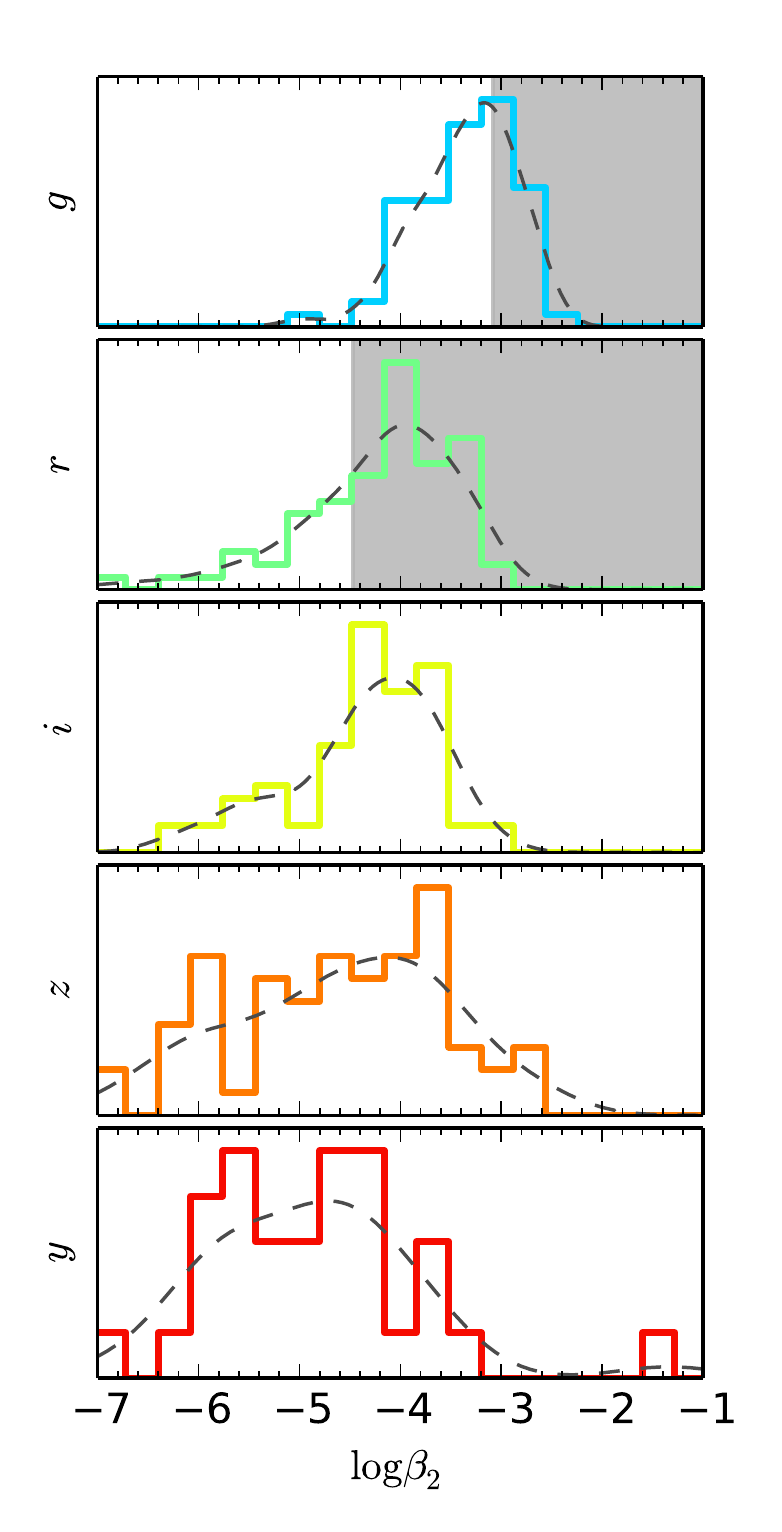}
\caption{\label{fig:betakde}Investigation of a potential SN~IIL sub-population.  The histograms display the distribution of mean posterior values for the $\beta_2$ plateau decline rate parameter in our fitted light curve models in each photometric band.  Only objects classified nominally as SNe~IIP (i.e. not classified as SNe~IIn or IIb, see Section~\ref{sec:mod:class}), and only objects with well constrained $\beta_2$ posteriors (with variance $\leq80\%$ of the prior variance), are included.  The dashed lines display the Gaussian kernel density estimate models for the distributions.  The $g$-band shaded region shows the approximate range for objects that would be classified as SNe~IIL under the \cite{Barbon79} $B$-band criterion (with more positive $\beta_2$ values reflecting faster decline rates), and the $r$-band shaded region shows the \cite{Li11} $R$-band criterion.}

\end{figure}

We take an empirical approach to searching for evidence of a fast-declining mode in the sample of PS1 objects classified nominally as SNe~IIP (see Section~\ref{sec:mod:class}).  Figure~\ref{fig:betakde} shows the distribution of mean posterior values for the $\beta_2$ plateau decline rate parameter in our fitted light curve models, among objects with significant constraints on the plateau phase decay behavior.  We employ Gaussian kernel density estimation (KDE) to estimate the underlying form of the decline rate distribution from the observed data.  We select a KDE bandwidth factor of \BETAkdefactor\ in $r$-band (and similar in other bands) based on the number of objects in the well constrained sample, following \cite{Scott92}.  Inspection of the KDE models in Figure~\ref{fig:betakde} does not indicate the presence of any fast-declining modes, e.g. within the shaded regions corresponding to the \cite{Barbon79} and \cite{Li11} criteria.  Given the limited sample size, this test would not be sensitive to the existence of a small SN~IIL sub-population if they are subsumed within the fast-declining tail of a more numerous SN~IIP population.  However, a significant sub-population ($\sim20\%$) as proposed by previous authors should emerge if present.

We therefore conclude that empirical evidence for a distinct SN~IIL sub-population is not present within the PS1 SN~II dataset, and hereafter we do not distinguish between SNe~IIL and IIP.  Discounting the existence of an SN~IIL subclass differs from the interpretation applied in many past studies, but does not disagree quantitatively with past results.  \cite{Li11} adopted an SN~IIL classification scheme defined arbitrarily by previous authors \citep{Barbon79} for the purpose of estimating volumetric rates.  \cite{Patat94} applied a similar criterion, then tested for a discontinuity between the properties of this sub-population and the ``normal'' SN~IIP population.  They reported a negative result.  \cite{Arcavi12} identified a slow declining subset of objects in their sample, but did not quantitatively test for a discontinuity.  Moreover, we note that our sample, consisting of \NIItotal\ SNe~II in total, is $\sim40\%$ larger than the \cite{Li11} SN~II sample and $\sim5$~times larger than the \cite{Arcavi12} sample.  Using a similarly sized sample, \cite{Anderson14phot} have reported independent results in agreement with the continuum of decline rates we report here.

Figure~\ref{fig:stackLC} further illustrates this result, showing our SN~IIP $r$-band photometry in comparison to the averaged unfiltered SN~IIP and SN~IIL light curves of \cite{Li11}.  Both their templates fall within the photometric range of our observed sample, with their SN~IIP template falling near the $+1\sigma$ range of our light curves and their SN~IIL template falling near the median of our light curves.  Note that their photometry was collected with an unfiltered photometric system, which \cite{Li11} compare to $R$ band.  This would suggest that their response function includes redder wavelengths than our $r$-band photometry.  This may partially explain the slower decline rates in their sample, as redder filters (e.g. $izy$-bands) exhibit more gradual SN~IIP decline rates in the plateau phase.  Regardless, this comparison illustrates that the long-recognized range of decline rates among SNe with broad H$\alpha$ features falls within the observed, continuous range of decline behavior exhibited by objects classified as SNe~IIP in our sample, with no evidence for a second, fast-declining mode in the population.

\subsection{Plateau duration distribution}
\label{sec:res:tp}

Recent observational studies have suggested that the plateau durations of SNe~IIP are tightly distributed around $\sim100$~days \citep{Poznanski09,Arcavi12}, in stark contrast to theoretical predictions that they should vary from $\sim80-200$~days given the expected range of progenitor properties including mass, energy, and radius (see e.g. \citealt{Kasen09}).  \cite{Poznanski13} has interpreted this as evidence that the joint parameter space of progenitor properties must be tightly constrained, with the explosion energy scaling precisely with the cube of the mass to produce a roughly constant plateau duration.

\begin{figure}
\plotone{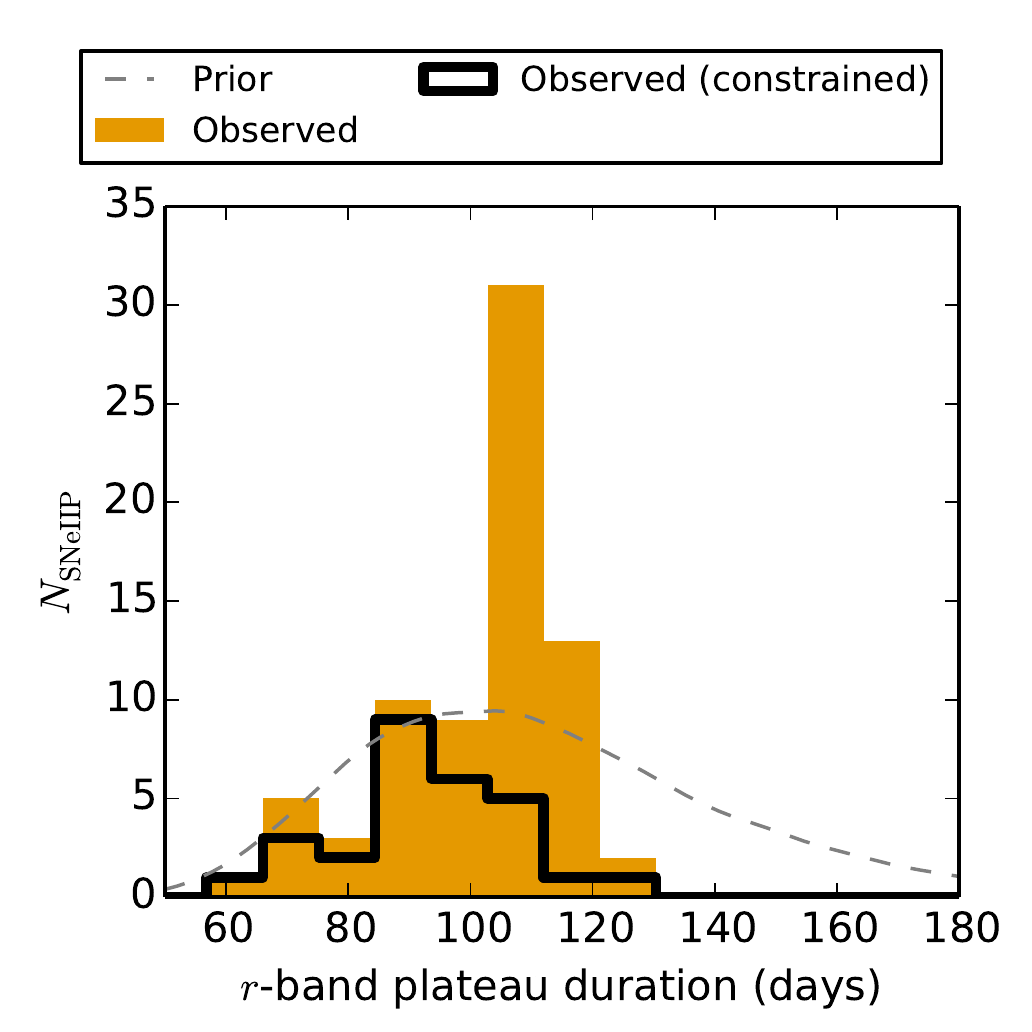}
\caption{\label{fig:tpfdist}The distribution of $r$-band plateau durations ($t_p + t_2$) among the SNe~IIP in our sample, as derived from the posterior medians of our Bayesian light curve fits.  The solid line shows the subset of objects with well-constrained marginalized posterior uncertainties ($<15$~days at $1\sigma$; $N=\tpfN$).  The dashed line shows the lognormal mixture prior on the plateau duration.}
\end{figure}

We quantitatively assess the plateau duration distribution of our SNe~IIP sample.  In agreement with \cite{Arcavi12}, \cite{Poznanski13}, and \cite{Faran14}, the $r$-band plateau duration distribution of our sample (Figure~\ref{fig:tpfdist}) is peaked around $\sim90-100$~days, with median value of \tpfM~days and $1\sigma$ variation of \tpfrstd~days.  However, the full observed range of plateau durations is significantly larger, $\tpfmin-\tpfmax$~days.  Note that we do not include in this distribution poorly-constrained light curves, where the marginalized posterior uncertainty is $>80\%$ of the prior variance ($\tpfrstdprior$~days at $1\sigma$).  For those poorly constrained objects, as Figure~\ref{fig:tpfdist} illustrates, the posterior is typically peaked around the mean of the prior distribution.

Our PS1 sample therefore suggests that the observed variation in SN~IIP plateau durations is not as highly constrained as suggested by \cite{Poznanski09} and \cite{Arcavi12}.  However, we do not detect objects with plateaus as long lived as the least energetic objects produced in the model grid of \cite{Kasen09} (e.g. $\sim200$~days for kinetic explosion energy of $0.3~B$ at infinity).  However, all wide field surveys will be insensitive to identifying the longest duration objects for two primary reasons.  First, such low energy explosions will have correspondingly low luminosities ($L_{50}\sim41.8$, similar to the least luminous objects in our sample; see Table~\ref{tab:ppars}) and therefore be subject to Malmquist bias.  Second, because the length of the observing seasons of each PS1 MD field (typically $\lesssim160$~days) are similar to the SN~IIP plateau duration, there will be an observational bias towards measuring well-constrained plateau durations only for short-duration objects.  We therefore interpret the results presented here as a lower limit on the true variation in the SN~IIP plateau duration distribution.  In a companion paper, \cite{Sanders14Unsup}, we use a hierarchical Bayesian light curve fitting methodology to assess the influence of these biases on our inferred plateau duration distribution, and to estimate the underlying plateau duration distribution.

\subsection{Line of sight extinction}
\label{sec:prog:ext}

\begin{figure}
\plotone{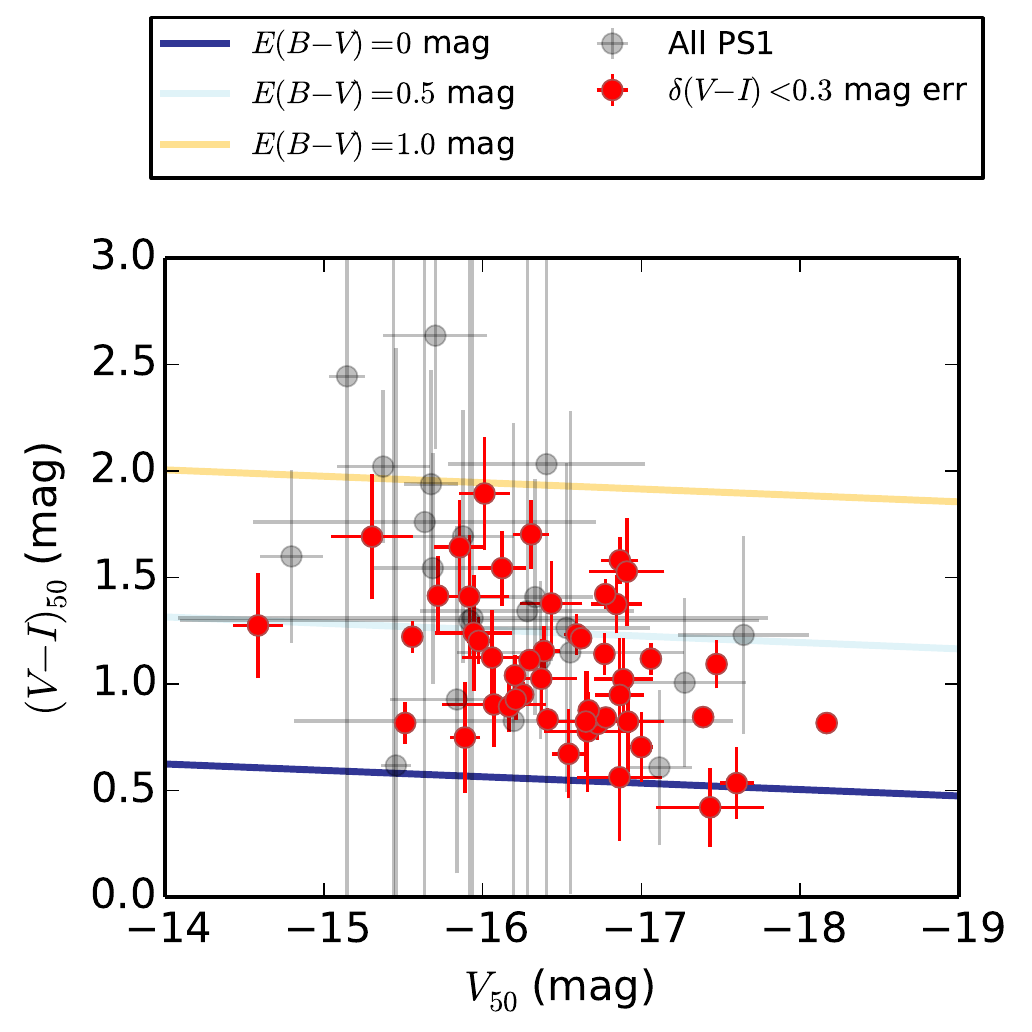}
\caption{\label{fig:VIext}Illustration of $(V-I)$ color excess method for extinction estimation.  The points show the $V$ magnitude and $(V-I)$ color (in the Vega magnitude system) at $50$~days for objects from our SN~IIP sample, with error bars representing $1\sigma$ variation in the posterior predictive luminosity distribution for each object.  Objects with strong posterior color constraints are highlighted in red.  The solid lines show theoretical color-magnitude relations for SNe~IIP with different levels of reddening, based on the fiducial relation of \cite{Kasen09}.}
\end{figure}

We correct for total extinction (Galactic and host environment) using the $(V-I)$ color excess method of \cite{Nugent06} as revised by \cite{Kasen09}---see Figure~\ref{fig:VIext}.  We calculate $E(V-I)$ by comparing to the theoretical baseline $(V-I)$ color\footnote{We convert our $griz$ photometry to the Landolt and Vega systems using a color-based $S$ correction obtained by integrating over the temporally nearest available UVOIR spectrum of SN~1999em, the $\sim+37$~day spectrum from \cite{Leonard02}.  We find that $s$-corrections based on the spectrophotometric templates applied in Section~\ref{ssec:Kcor} would introduce a systematic discrepancy in the $(I-i)$ color relative to SN~IIP corrections in past works (e.g. \citealt{Dandrea10}) of $\sim0.4$~mag, due to differences in the equivalent width of the NIR Ca~II emission feature.  We estimate a $\sim0.1$~mag systematic uncertainty may remain in our $I$-band magnitude estimates due to variation in this feature.  We apply the AB offset of \cite{Blanton07}.} for the given $V$-band luminosity (Equation~15 of \citealt{Kasen09}), and convert to extinction in each $griz$ band using the extinction curve of \cite{cardelli89} and assuming $R_V=3.1$.  We apply a physically-motivated uniform prior enforcing $E(V-I)\geq0$~mag.  The median uncertainty for the extinction estimate of an individual SN is \EBVsigmed~mag.

We find extinctions varying from $\EBVmin-\EBVmax$~mag, with median value of $\langle A_V\rangle = \EBVmed$~mag.  In this distribution we have excluded objects with extinction uncertainties $\geq2$~mag, affecting \EBVsigNexc\ objects.  This median is somewhat higher than the mean extinction reported by \cite{Smartt09MNRAS} for their sample of SNe~IIP with progenitor imaging, $\langle A_V\rangle = 0.7\pm1.1$~mag.  The discrepancy may result from a combination of differences in methodology (the estimates from \citealt{Smartt09MNRAS} are derived from a heterogeneous set of sources and techniques) and sample selection (given that our PS1 transient search operates in red bands, particularly $i$ and $z$, it may be relatively insensitive to selection against high extinction objects).  
Both observations and theory show that there is significant scatter around the fiducial $(V-I)$ color, indicating intrinsic color variation not accounted for in our color excess-based extinction estimate.  Moreover, the color excess method may be particularly vulnerable to under estimates for low-metallicity progenitors where decreased iron line blanketing would reduce suppression of blue flux.  
We note that \cite{Faran14} have recently shown that both photometric and spectroscopic SN~IIP extinction correction prescriptions have limited effectiveness (as measured by their ability to reduce the scatter in observed colors), and exhibit systematic offsets relative to each other.

\subsection{Peak magnitude distribution}
\label{sec:res:mpeak}

\begin{figure*}
\plotone{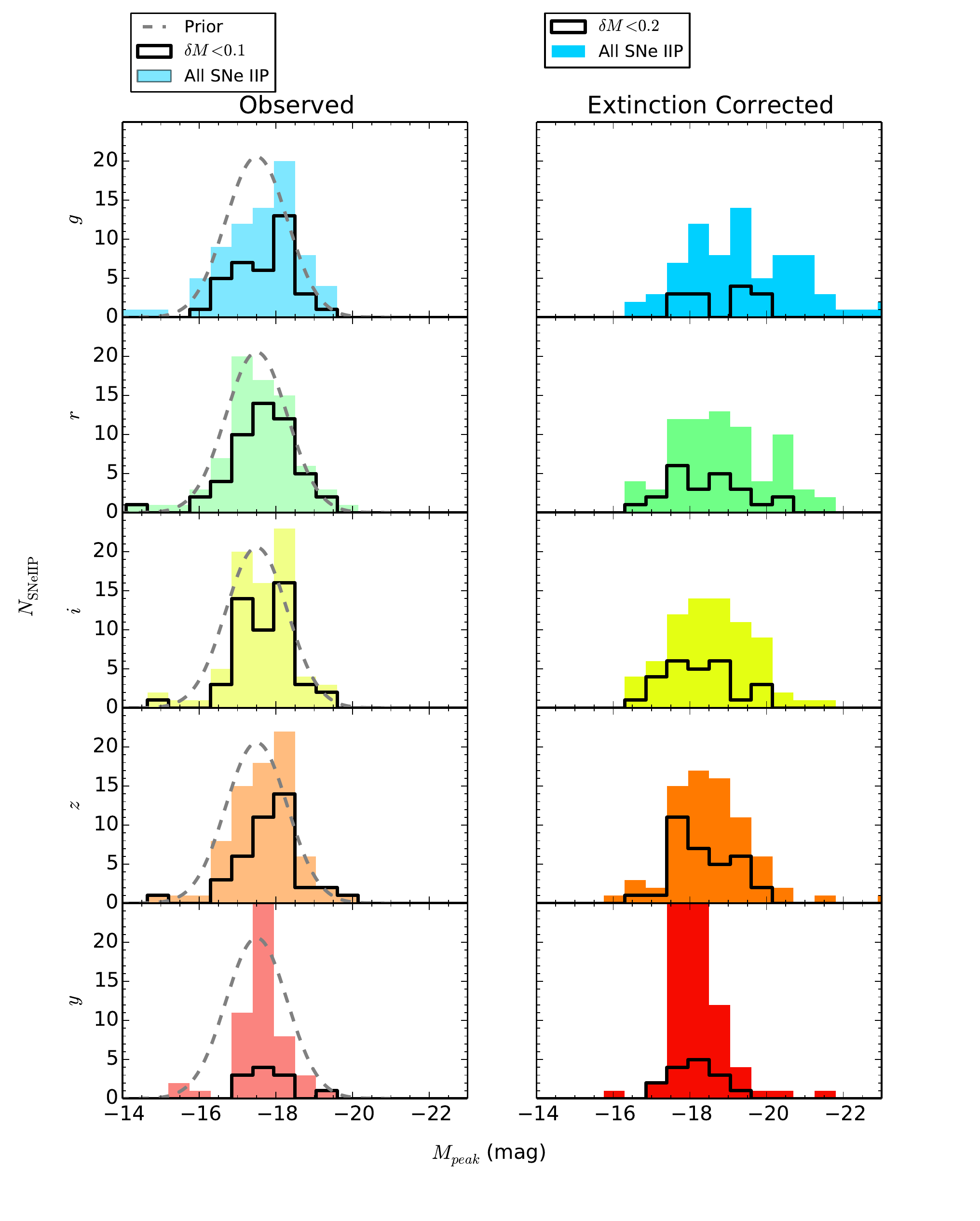}
\caption{\label{fig:mpeakdist}The peak absolute magnitude distribution among the SNe~IIP in our PS1 sample, as derived from the posterior medians of our Bayesian light curve fits, without (left) and with (right) correction for extinction (see Section~\ref{sec:prog:ext}).  The solid line shows the subset of objects with well-constrained marginalized posteriors ($<0.1$~mag at $1\sigma$ without extinction correction, $<0.2$~mag with correction).  The dashed line shows the prior distribution discussed in Section~\ref{sec:mod:fit}.  All data are $K$-corrected.}
\end{figure*}

We show the peak absolute magnitude distribution for our PS1 SN~IIP sample in each photometric filter in Figure~\ref{fig:mpeakdist}.  We correct for extinction using the $(V-I)$ color excess method, as described in Section~\ref{sec:prog:ext}.  Statistics of these distributions are given in Table~\ref{tab:absmagdist}.  In this table and hereafter, we include only objects with well constrained posterior distributions: $\delta M_{peak}<0.1$~mag at $1\sigma$ without extinction correction, and $\delta M_{peak}<0.2$~mag with correction.  The median uncertainty for the absolute magnitude measurement ($1\sigma$ posterior width) is $\delta M_{peak}$ of $\mpeakdmedr$~mag in $r$~band without extinction correction, and $\mpeakCdmedr$~mag after correction.

The observed peak magnitude distribution agrees well with the $R$ band luminosity function of \cite{Li11}, reflected in the prior distribution shown in Figure~\ref{fig:mpeakdist} (see also Section~\ref{sec:mod:fit} and Figure~\ref{fig:mpprior}).  The comparison is somewhat limited, as \cite{Li11} examined a different filter ($R$) and did not correct for extinction (although the objects in their sample should have low extinction).  Note that this similarity suggests agreement of the observed distribution of peak magnitudes between our surveys.  If the prior were simply dominating our inferred peak magnitudes, the posterior medians would all be centered at the prior mean ($M_{peak}=-17.5$).  This effect can be seen in the $y$-band peak magnitude distribution shown in Figure~\ref{fig:mpeakdist}, where prior information does dominate for a substantial fraction of objects.  Extinction correction shifts the median of the distribution by $\sim0.6$~mag in $g$-band, and only $\sim0.1$~mag in $z$-band.  We regard the $y$-band distribution as least reliable due to the small number of objects ($N=\mpeakCNy$) that pass our posterior width cut, and therefore do not include them in Table~\ref{tab:absmagdist}.

The extinction corrected peak magnitude distribution spans a $1\sigma$ range of $\mpeakCUr$ to $\mpeakCDr$~mag in $r$-band, indicating a factor of $\sim3$ spread in the luminosity distribution of these SNe.  This wide range in SN properties has important implications for the mass distribution of their progenitor stars, which we will model and discuss in Section~\ref{sec:prog:mass}.  \cite{Richardson14} found a $B$-band volume limited peak magnitude distribution with mean and standard deviation of $M_B\sim-17\pm1$ among 74 SNe~IIP from the literature, $\sim1.5$~mag dimmer than the $g$-band mean peak magnitude we report.  The discrepancy in the mean of the absolute magnitude distributions between our study and theirs is likely due to a combination of sample selection effects and extinction correction methodology.  \cite{Richardson14} have considered SNe~IIL separately from SNe~IIP, which systematically lowers their SN~IIP peak magnitude distribution relative to ours (see Section~\ref{sec:res:IIL}).  Morever, \cite{Richardson14} have applied a statistical correction for extinction, assuming a mean value of $A_B\sim0.3$~mag for each core-collapse SN, which is typically $\sim1$~mag lower than the extinction correction we measure among our SNe~IIP with the $(V-I)$ color excess method (Section~\ref{sec:prog:ext}).

\begin{deluxetable}{llrrrlrrr}
\tablecaption{SN~IIP Absolute Magnitude Distribution Statistics\label{tab:absmagdist}}
\tabletypesize{\scriptsize}
\tablehead{
     & \multicolumn{4}{c}{{Observed}} & \multicolumn{4}{c}{{Extinction Corrected}} \\
    \colhead{F}      & N & 16th & 50th & 84th & N & 16th & 50th & 84th 
	   }
\startdata
$g$ & \mpeakNg & \mpeakUg & \mpeakMg & \mpeakDg & \mpeakCNg & \mpeakCUg & \mpeakCMg & \mpeakCDg \\
$r$ & \mpeakNr & \mpeakUr & \mpeakMr & \mpeakDr & \mpeakCNr & \mpeakCUr & \mpeakCMr & \mpeakCDr \\
$i$ & \mpeakNi & \mpeakUi & \mpeakMi & \mpeakDi & \mpeakCNi & \mpeakCUi & \mpeakCMi & \mpeakCDi \\
$z$ & \mpeakNz & \mpeakUz & \mpeakMz & \mpeakDz & \mpeakCNz & \mpeakCUz & \mpeakCMz & \mpeakCDz
\enddata
\tablecomments{The $[16,50,84]$th columns correspond to percentile values within the SN~IIP absolute, $K$-corrected peak magnitude distribution, without (``observed'') and with extinction correction (Section~\ref{sec:prog:ext}).  Only PS1~SNe~IIP with low posterior uncertainty in their peak magnitudes are included ($\delta M<0.1$~mag at $1\sigma$ without extinction correction, $\delta M<0.2$~mag with); the total number of objects in each distribution (N) is given.  Each row corresponds to a PS1 photometric filter (F).}
\end{deluxetable}

\subsection{Decline rate--peak magnitude relation}
\label{sec:res:pcor}

\begin{figure*}
\plotone{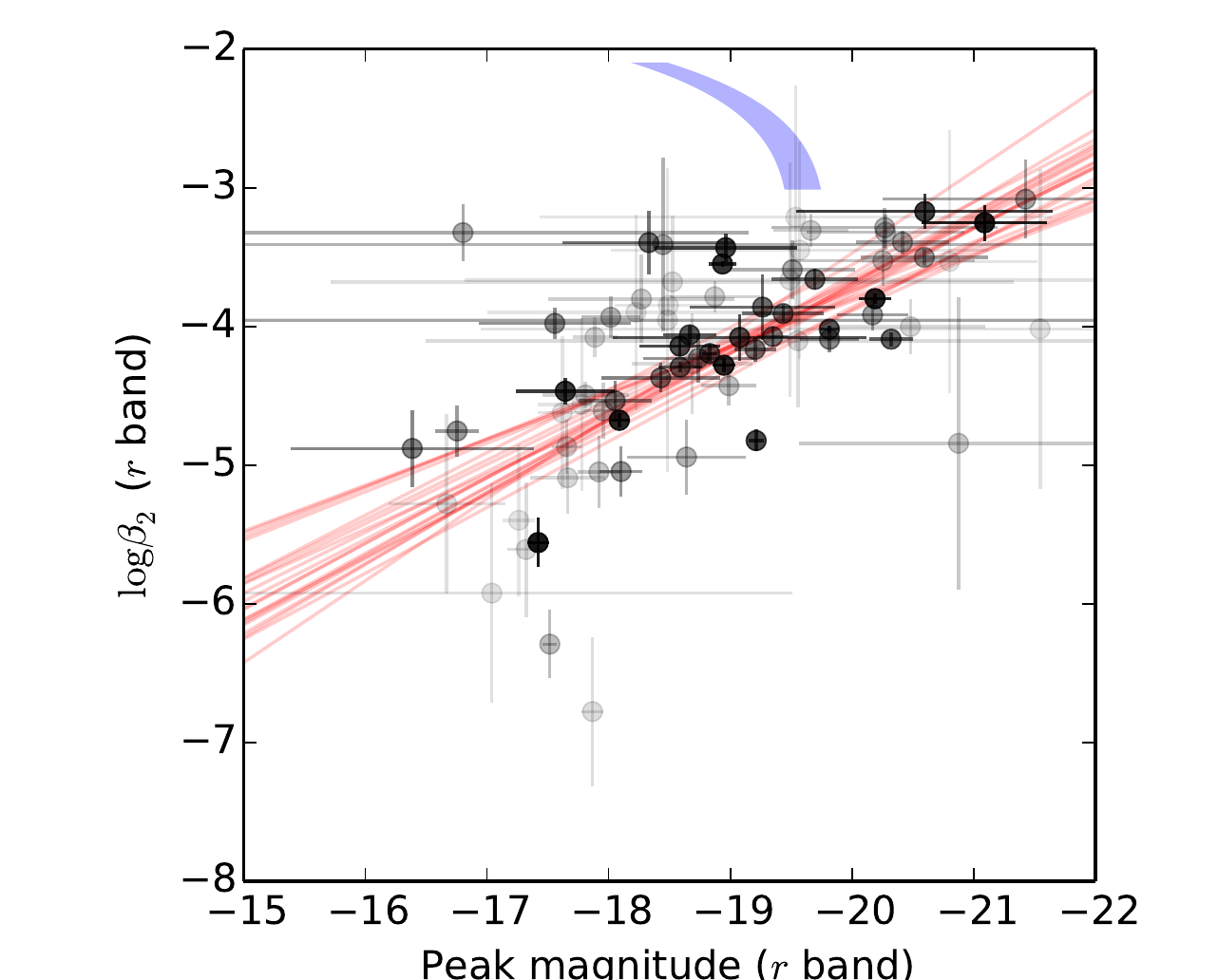}
\caption{\label{fig:phaseMprBeta2}Relation between the plateau phase decay rate ($\beta_2$, $r$-band) and the extinction corrected $r$-band peak magnitude ($M_{peak}$) among SNe~IIP in the PS1 sample.  The red lines show posterior samples of linear fits to the parameter values.  The opacity of the points is drawn in proportion to the inverse of their posterior variance.  The blue band shows the SN~Ia width--luminosity relation of \cite{Phillips99}, assuming $M^{max,Ia}_r=-19.45$~mag and a dispersion of $0.15$~mag.}
\end{figure*}

\begin{figure}
\plotone{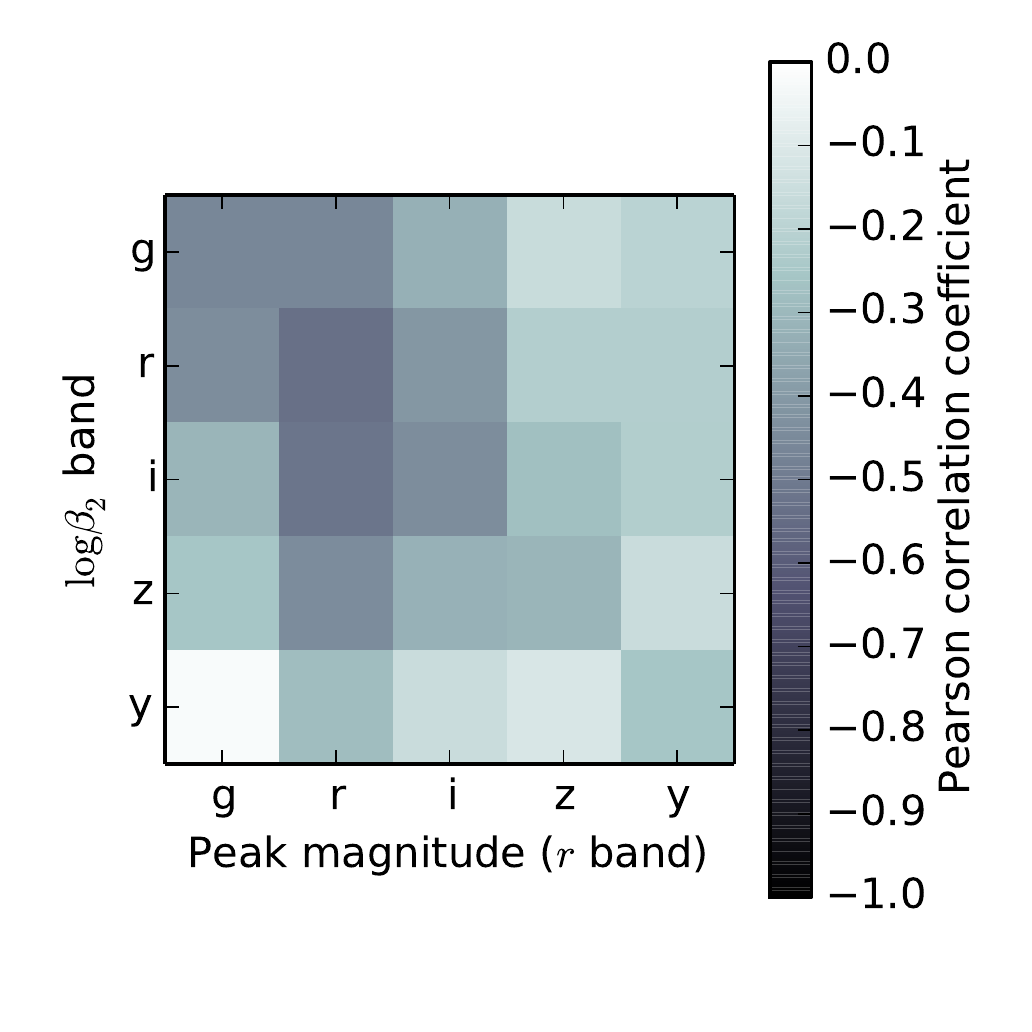}
\caption{\label{fig:phaseMprBeta2bands}Comparison of $\beta_2-M_{peak}$ (extinction corrected) relation for different photometric filter sets.  The shading represents the strength of the correlation between the two variables in each pair of photometric filters, as measured by the Pearson correlation coefficient (key at right).}
\end{figure}

Upon investigation of the joint posterior parameter distributions among our fitted SN~IIP light curves, we identified a highly significant relation between the plateau phase decline rate ($\beta_2$) and the extinction corrected peak magnitude ($M_{peak}$) of the SNe.  Figure~\ref{fig:phaseMprBeta2} illustrates that these parameters are highly correlated in $r$-band, with Pearson correlation coefficient of $\betatwoMpCP$ ($p$-value of $\betatwoMpCPp$).  We have evaluated the strength of the correlation between these parameters in all combinations of photometric filters available from the PS1 dataset (Figure~\ref{fig:phaseMprBeta2bands}) and identified this relation to be significant in every band (except $y$, where the PS1 photometry is limited), though it is strongest in the $r$ and $i$-bands.

In the $r$-band, the relation between these parameters is best fit with the relation:

\begin{align}
\log \beta_2[r] & = (\betatwoMpCBFamed \pm \betatwoMpCBFastd) + M_{peak}[r] ~ (\betatwoMpCBFbmed \pm \betatwoMpCBFbstd)
\end{align}

\noindent This linear fit was obtained using a Bayesian methodology accounting for the two dimensional covariance of the $M_{peak}$ and $\beta_2$ posterior distributions for each SN, and modeling the relationship between these values with an intrinsic scatter $V$ orthogonal to the axis of the fit line \citep{HoggModel}.  

The $[16,50,84]$th percentile values of the intrinsic scatter parameter, projected to the peak magnitude axis, are $[\betatwoMpCBFVdown,\betatwoMpCBFVmed,\betatwoMpCBFVup]$~mag.  No samples among the $20,000$ returned from our MCMC fit of this model have positive slopes for the linear fit, supporting the low $p$-value reported from the Pearson correlation coefficient test above.  When we perform the same analysis on the observed absolute peak magnitudes, not corrected for extinction, we find a similar trend, but with somewhat decreased significance ($p$-value of $\betatwoMpPp$ and projected intrinsic scatter median $V=\betatwoMpBFVmed$~mag).  This result suggests that application of the color-excess based extinction estimates provides measurable success in recovering the intrinsic distribution of light curve properties, despite the shortcomings described in Section~\ref{sec:prog:ext}.  Considering the $V$-band data of \cite{Anderson14phot}\footnote{Here we exclude SN 2008bp, which \cite{Anderson14phot} identify as an outlier and special case, and we transform their $s_2$ parameters to our $\beta_2$ given that $\log\beta_2=\log(s_2/108.6)$.} and applying the same Bayesian linear fitting methodology yields a median intrinsic scatter of $V=\AndersonbetatwoMpCscatter$~mag.  The larger intrinsic scatter indicated by the \cite{Anderson14phot} data results in part from the much smaller uncertainties in the light curve parameters of the nearby SNe they study.  This discrepancy may imply that the uncertainties in our light curve parameters are somewhat overestimated, or that theirs are somewhat underestimated.

In past studies, divisions of SN~IIP samples into ``normal'' SN~IIP and SN~IIL (see Section~\ref{sec:res:IIL}) sub-classes may have obscured this relation by reducing the dynamic range available for correlations against the decline rate.  \cite{Anderson14phot} have recently reported an independent discovery of this relation using a similarly-sized sample of $V$-band light curves from a variety of past SN~searches.  This independent discovery, and the multi-band manifestation of the relation that we explore here, underscores its physical significance.  This result echoes past conclusions that ``SNe~IIL'' are on average more luminous than more slowly declining SNe~IIP (e.g. \citealt{Patat94,Li11}).

The existence of a decline rate--peak magnitude relation for SNe~IIP is reminiscent of the well known width--luminosity relation for SNe~Ia \citep[e.g.][]{Phillips99}.  However, while brighter SNe~Ia have  more slowly declining light curves, brighter SNe~IIP instead have faster declining light curves (Figure~\ref{fig:phaseMprBeta2}).  Moreover, the SN~IIP relation has significant intrinsic scatter: $V\sim\betatwoMpCscatter$~mag.  That the SN~IIP decline rate--peak magnitude relation is recoverable at all is testament to the wide range of variation in their plateau phase decline rates, which ranges a factor of $\sim 20$.

The SN~Ia decline rate--peak magnitude relation is interpreted as evidence for a fundamental plane or single-parameter family among SNe~Ia light curves, and a similarly sparse dimensionality may apply to SNe~IIP.  Several authors have argued for a single parameter source of variation among SNe~IIP light curves \citep{Hamuy03,Poznanski13}, particularly with respect to the luminosity--velocity relation applied in the standardizable candle method \citep{Hamuy02SCM,Olivares08,Dandrea10}.  Interpreted together with the results of Section~\ref{sec:res:IIL}, this would seem to point to the explosion energy, likely determined by the progenitor initial mass through its influence on the mass of the hydrogen envelope at the time of explosion, as the predominant single parameter continuously determining the fundamental observational properties of the explosions of hydrogen-rich red supergiants, as \cite{Anderson14phot} also concluded.  However, the presence of significant scatter around the SN~IIP decline rate--peak magnitude relation presented here suggests additional correlations with secondary factors, as have emerged among SNe~Ia \citep{Wang09,FSK11}.  Relevant secondary characteristics of the progenitor stars are likely to include metallicity and rotation velocity.  

In general, more massive H-rich envelopes should have longer radiative diffusion times ($t_{\rm{diff}}\propto M / R$), suggesting slower light curve evolution and lower peak magnitudes.  However, \cite{Poznanski13} have found an empirical linear relation between the progenitor initial mass of SNe~IIP and their expansion velocities in the plateau phase.  With these two factors in opposition, this would suggest that the diffusion time, and perhaps the $\beta_2$~parameter, should be nearly constant among SNe~IIP.  This discrepancy could result from nonlinearity in the relationship between progenitor initial mass and the mass of the hydrogen envelope at the time of explosion or variation in the velocity evolution between SNe~IIP.  Alternatively, it may challenge the robustness of the linear mass--velocity relation constructed by \cite{Poznanski13}.

\section{Progenitor Modeling}
\label{sec:prog}

We compare the light curve models for each SN~IIP in our sample to relevant theoretical models to estimate the physical properties of their progenitor stars.

\subsection{Bolometric luminosity}
\label{sec:prog:BC}
\label{sec:res:bolL}

For the purpose of comparing to the full-wavelength spectrophotometric models of \cite{Kasen09}, we estimate the pseudo-bolometric (OIR) luminosity of each SN by integrating over the multi-band ($grizy$) light curve model described in Section~\ref{sec:mod:fit}.  We apply a correction to the pseudo-bolometric light curves to, primarily, account for unobserved IR flux and some UV emission.  We calculate this correction ($\rm{pBC}$) by integrating a SN~IIP spectral template from a similar epoch ($t=+61$~days; \citealt{Gilliland99,Baron04}) over the $'griz'$ bands, obtaining $\rm{pBC}=0.38$~mag.  By using a fixed spectral template, we neglect any intrinsic variation in the SN~IIP spectral properties.  The modeling in \cite{Kasen09} suggests that any intrinsic color variation among SNe~IIP is small, with the bolometric correction estimated to vary by $\sim0.1$~mag based on differences in progenitor metallicity and initial mass (their Figure~14).

\subsection{Ejected Ni mass}
\label{sec:res:decay}

\begin{figure}
\plotone{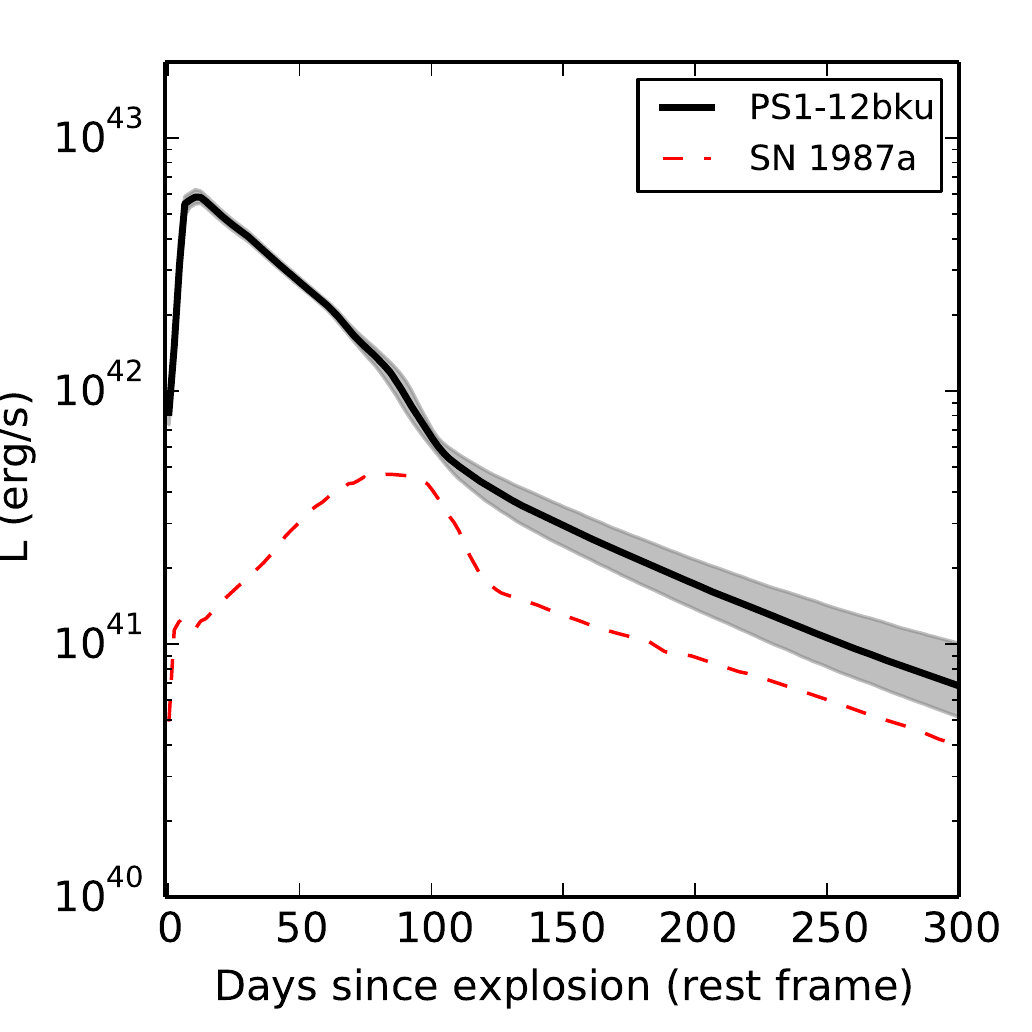}
\caption{\label{fig:qbolo}Illustration of ejected nickel mass ($M_{\rm{Ni}}$) measurement methodology.  The pseudo-bolometric (optical-infrared) light curve for {\PSOfourtwozerothreeninethree} (black solid line) is shown in comparison to the late time bolometric light curve of SN~1987A (red dashed line; from \citealt{Pastorello04} and references therein).  The $1\sigma$ uncertainty in the bolometric luminosity as derived from the multi-band light curve model is illustrated with the shaded region, including the uncertainty in the explosion epoch.}
\end{figure}

We estimate the ejected mass of radioactive $^{56}$Ni ($M_{\rm{Ni}}$) by comparison to the late-time light curve of SN~1987A (see Figure~\ref{fig:qbolo}), following the method of \cite{Pastorello04}.  We compare our pseudo-bolometric light curves (Section~\ref{sec:res:bolL}) to the SN~1987A light curve using the late-time ($>125$~days) rest frame epoch where the model is most tightly constrained and assuming complete gamma-ray trapping, as in \cite{Pastorello04}; but see also the discussion of gamma-ray trapping in \cite{Anderson14phot}.

\begin{figure}
\plotone{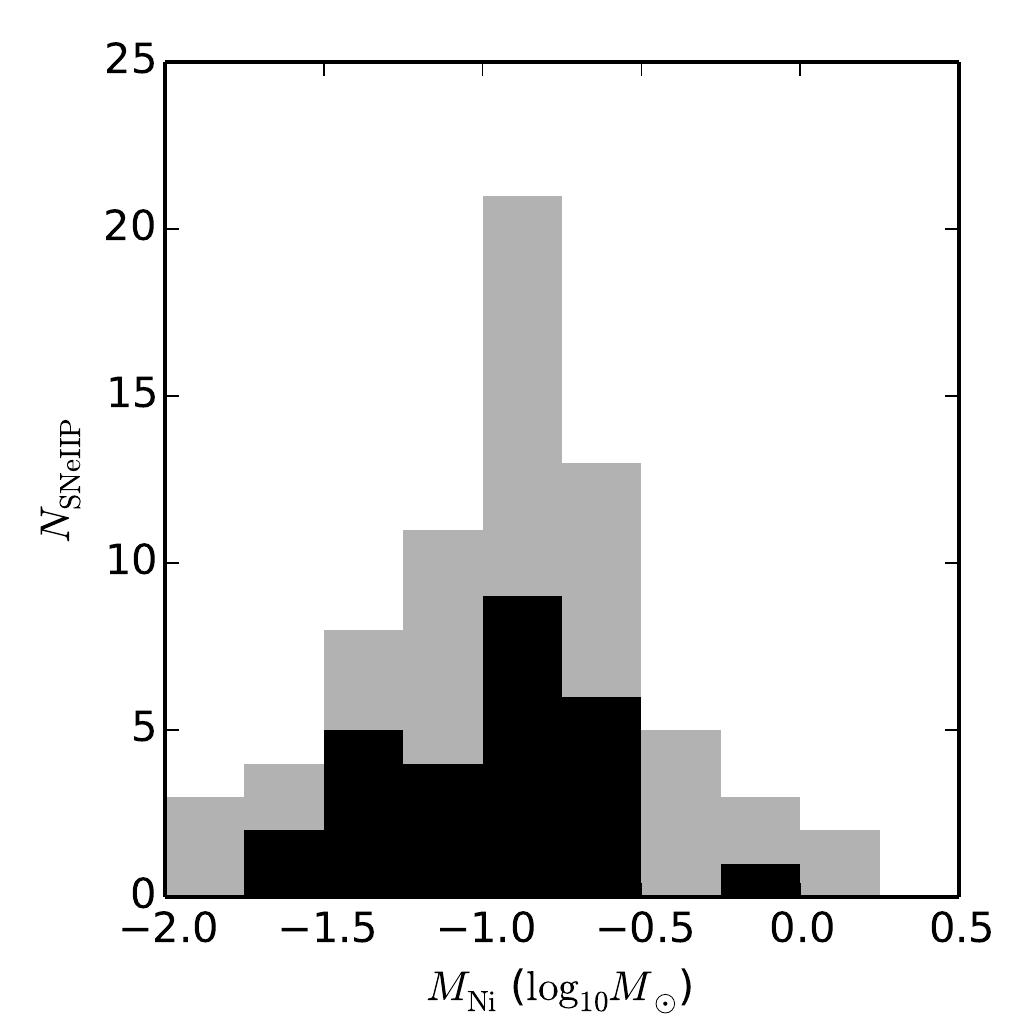}
\caption{\label{fig:MNidist}The distribution of ejected nickel mass ($M_{\rm{Ni}}$) among the SNe~IIP in our sample, as derived from the late time bolometric luminosity.  The gray shaded area shows the distribution estimated among all objects in our sample, and the black shaded region shows only objects with photometry constraining the light curve evolution in the radioactive decay phase (based on the variance in the $\beta_{dC}$ posterior; see text).}
\end{figure}

The distribution of $M_{\rm{Ni}}$ among our PS1 SNe~IIP has $[16,50,84]$th percentile values of $[\MNiD,\MNiM,\MNiU]~\rm{M}_\odot$ (see Figure~\ref{fig:MNidist}).  We include in this distribution only objects with photometry to directly constrain their light curve evolution in the radioactive decay dominated phase.  We identify such objects by looking at the posterior variance in the $\beta_{dN}$ parameter, which is $\delta\beta_{dN}=1$~dex at $1\sigma$ in cases where the prior alone sets the parameter value.  We therefore adopt a cut at $\delta\beta_{dN}<0.9$~dex to exclude poorly constrained Ni masses.  We note that, among the entire sample, the typical uncertainty in $M_{\rm{Ni}}$ is fairly large (median of $\MNidUdex$~dex) due to the relatively poor photometric coverage of the late phases of SNe in the sample.  Among the excluded objects, some have unphysically large median estimates for the ejected $^{56}$Ni mass ($M_{\rm{Ni}}>1~\rm{M}_\odot$); the high uncertainty in these estimates is appropriately reflected in the confidence intervals quoted in Table~\ref{tab:ppars}.

Our results suggest a factor of $\sim2$ smaller $M_{\rm{Ni}}$ for most SNe~IIP in comparison to typical SNe~Ib and Ic, which have $\langle M_{\rm{Ni}}\rangle \approx 0.20\pm0.16~\rm{M}_\odot$ \citep{Drout11}.  This is consistent with the expectation that SNe~Ib and Ic are produced by more massive stellar progenitors than SNe~IIP.  Our measured $M_{\rm{Ni}}$ distribution is similar to that reported by \cite{Nadyozhin03}, who derive $M_{\rm{Ni}}$ ranging from $0.03-0.4~\rm{M}_\odot$ for a set of 11 SNe~IIP from the literature, with a mean value of $0.10~\rm{M}_\odot$.  

\subsection{Comparison to hydrodynamic models}
\label{sec:res:prog}

We derive physical parameters (including progenitor initial mass, radius, and explosion energy) for the progenitor stars of our SN~IIP sample by comparison to the modeling of \cite{Kasen09}.  In general, more massive progenitor stars produce plateaus that are brighter and of longer duration.  There is a secondary contribution from $M_{\rm{Ni}}$ which has the effect of increasing the plateau duration.  \cite{Kasen09} provide scaling relations for the observable parameters $L_{50}$ (bolometric luminosity at $50$~days after explosion) and $t_{p,0}$ (plateau phase duration, corrected for extension by ejected $^{56}$Ni) as a function of $E_{51}$ (total explosion energy in units of $10^{51}$ ergs) and $M_{in}$ (initial mass of progenitor star).  We note that there is a well known discrepancy between masses derived from hydrodynamic light curve modeling and direct progenitor searches, with progenitor searches yielding lower mass estimates \citep{Utrobin08,Maguire10}.  

We derive $t_{p,0}$ using their Equation~13, which requires an estimate for $M_{\rm{Ni}}$ (see Section~\ref{sec:res:decay}), $R_0$ (the progenitor radius at the time of explosion) and $M_{ej}$ (the total ejected mass).  Following the recommendations of \cite{Kasen09}, we calculate values for $R_0$ and $M_{ej}$ strictly as a function of $M_{in}$ by interpolating over the progenitor star model grid used in their study.   Solving this system of equations for every sample in the posterior distribution of the light curve model, we obtain estimates for $M_{in}$, $M_{ej}$, and $E_{51}$ for each SN~IIP in our sample.

\subsection{Progenitor initial mass inference}
\label{sec:prog:mass}

\begin{figure*}
\plotone{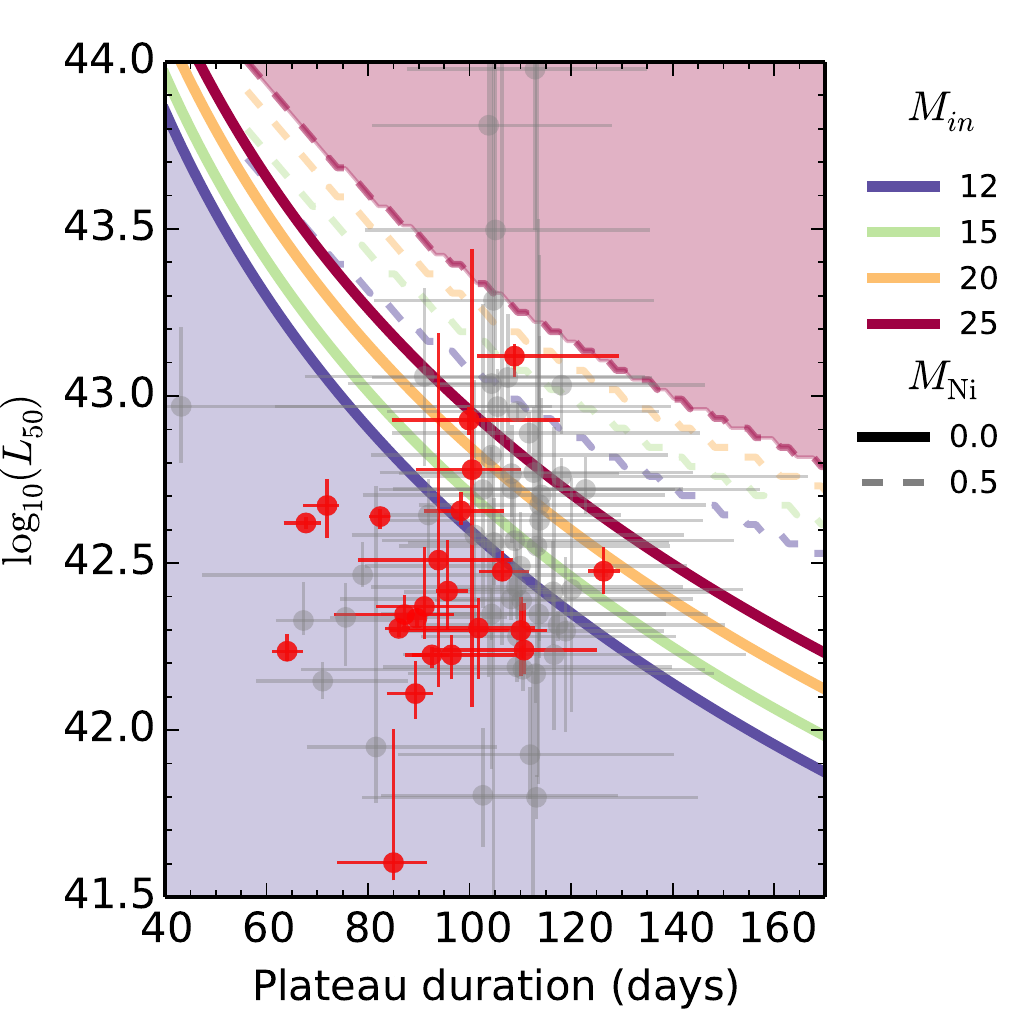}
\caption{\label{fig:ppar}Comparison of observed SNe~IIP light curve properties to hydrodynamic light curve models.  The points show the plateau duration ($t_p$+$t_2$, rest frame; not corrected for $M_{\rm{Ni}}$) of SNe~IIP from the PS1 sample versus their bolometric luminosity at 50~days after explosion ($L_{50}$).   Red points have well-constrained values for the plateau duration ($\delta t_p < 20\%$), while gray points have poorer constraints.  The solid lines show interpolated values for progenitors of different initial masses ($M_{\rm{in}}$; key at right) from the model grid of \cite{Kasen09}.  Different curves are plotted to illustrate the range of variation with the mass of ejected radioactive nickel ($M_{\rm{Ni}}=[0,0.50]~M_\odot$ for [solid, dashed] line styles, respectively).  The shaded curves illustrate physical parameter space not explored by the \cite{Kasen09} model grid, ostensibly corresponding to masses $<12$ or $>25~M_\odot$.}

\end{figure*}

In Figure~\ref{fig:ppar}, we compare the observed plateau duration and $L_{50}$ to model curves from \cite{Kasen09}.  This comparison illustrates that the observed variation in SN~IIP light curve phenomenology is significantly larger than the range explored by the \cite{Kasen09} model grid, preventing us from performing progenitor parameter estimation for a substantial number of the objects in our sample.  Given the observed $M_{\rm{Ni}}$ values measured for each individual light curve, \MinNoutofbounds\ of our SNe~IIP have light curve properties outside the region covered by the  \cite{Kasen09} models, and we can produce mathematically valid comparisons to model light curves from the \cite{Kasen09} grid for only \MinNinbounds\ SNe~IIP (see Table~\ref{tab:ppars}).  Even these comparisons typically yield explosion energies at or above the most energetic models in their grid ($E_{51}>5$).  A large number of objects in our sample have luminosities lower than the range of theoretical values for a given plateau duration (falling in the blue shaded region of Figure~\ref{fig:ppar}); accounting for \MinNoutofboundslow\ of the \MinNoutofbounds\ objects for which we do not report progenitor inferences.  

We therefore advocate for further theoretical exploration of the SN~IIP light curve parameter phase space through additional hydrodynamic modeling, in order to provide more robust model grids and scaling relations for comparison to observational surveys.  In particular, certain modeling systematics should be addressed to enable inference on objects occupying the low-luminosity and short plateau duration parameter space.  \cite{Dessart11b} have previously recognized a systematic discrepancy between SN~IIP model and observed light curves, and identified over-sized radii in model progenitor stars as the likely cause.  Hydrodynamic explosion simulations based on more advanced progenitor star models, taking into account important effects such as rotation on the stellar radius (e.g. \citealt{Georgy13}), are needed to overcome this bias.  Moreover, synthetic light curves for SNe with progenitor initial masses $<12~M_\odot$, down to the limit for core-collapse (e.g. $8~M_\odot$, see \citealt{Smartt09MNRAS} and references therein), and higher explosion energies should be produced.  Such low mass progenitors have already been explored in the context of extremely low luminosity SN~IIP like SN~1997D \citep{Chugai00}, and model light curves of this kind calculated with full self-consistency across the mass and energy spectrum are needed.

Additionally, other progenitor physical parameters will play important secondary roles in determining SN light curve properties and should be included in future simulations.  In particular, lower metallicity can serve to lower the plateau luminosity by decreasing the opacity of the H envelope, therefore lowering plateau luminosities, and by decreasing mass loss, causing longer lived H recombination and plateau durations \citep{Kasen09,Dessart13}.  \cite{Dessart13} have recently performed a theoretical survey of SN~IIP radiative properties using a grid of synthetic progenitor stars expanded in parameter scope with respect to the grid of \cite{Kasen09}, including core overshoot, rotation, mixing, and metallicity.  However, the \cite{Dessart13} grid was limited to a single initial progenitor mass of $15~M_\odot$.  An expanded survey of the kind performed by \cite{Dessart13}, to provide updated observationally-applicable scaling relations of the kind provided by \cite{Kasen09}, would be an invaluable tool for inference on the progenitor population of massive star explosions in the nascent untargeted survey era.  Photometric galaxy metallicity diagnostics like that presented in \cite{SandersLZC} can be used to observationally estimate metallicities for the progenitor stars of observed SNe to facilitate model comparison.

\subsection{Implications for the Red Supergiant Problem}
\label{sec:prog:RSGprob}

While theoretical stellar evolution modeling has historically predicted that stars undergoing core-collapse up to $\sim25~M_\odot$ should produce Type~IIP SNe (see e.g. \citealt{Eldridge08,Smartt09} and references therein), a variety of observational techniques have failed to identify SN~IIP progenitors with masses greater than $\sim16~M_\odot$.  The shape of the stellar initial mass function can at least help to explain this absence, with $\sim8~M_\odot$ progenitors being a factor of $~10-15$~times more common than $\sim25~M_\odot$ progenitors.  However, even accounting for IMF prescriptions of varying steepness, \cite{Smartt09} reported an upper limit for the SN~IIP progenitor mass of $16.5~M_\odot$ based on their direct progenitor detection search.  Similarly, in an X-ray analysis of SN~IIP progenitor mass loss rates,  \cite{Dwarkadas14} estimated a maximum mass loss rate of $\sim19~M_\odot$.  Black hole fallback formation \citep{Fryer99,Heger03} and / or progenitor rotation \citep{Georgy13,Groh13b} may provide a theoretical solution to this ``Red Supergiant Problem.''  

Our results have only indirect implications for the Red Supergiant Problem.  While our progenitor mass estimates (e.g. Figure~\ref{fig:ppar}) show no evidence for an absence of SN~IIP progenitor stars at the high mass end of the \cite{Kasen09} progenitor grid, our mass estimates are only as reliable as the calibration of the hydrodynamic modeling underpinning them.  As we have discussed, there are known systematic offsets between the masses inferred from hydrodynamic modeling and progenitor star photometry, and additional theoretical investigation of this discrepancy is needed to conclusively resolve the Red Supergiant Problem.

However, our observations and modeling suggest that there is no discontinuous distinction between Type~IIP and IIL SNe (Section~\ref{sec:res:IIL}).  Given the correlation between plateau phase decline rate and luminosity (Section~\ref{sec:res:pcor}), and the theoretical connection between luminosity and progenitor mass, the exclusion of luminous ``SNe~IIL'' could obscure the high-mass end of the SNe~IIP progenitor population.  Emphasizing the role of small number statistics in this regime, we note that \cite{Smartt09} report a progenitor mass upper limit that is already consistent with $20~M_\odot$ at $2\sigma$, and consistent with $>25~M_\odot$ at $2\sigma$ if only events with robust progenitor detections are included in the analysis.  \cite{Walmswell12} have recently considered the extinction due to the dust produced by red supergiant winds and revised the SN~IIP upper mass limit to $20^{+6}_{-3}~M_\odot$ (with $2\sigma$ uncertainty).  As the direct progenitor search technique has so far only been applied to a handful of objects, we look to future detections of the progenitors of luminous SNe~IIP to address the Red Supergiant Problem (see e.g. \citealt{Fraser10,EliasRosa11}).

\section{CONCLUSIONS}
\label{sec:conc}

We have assembled and studied the full sample of Type~IIP SN light curves from the \PS\ Medium Deep Survey, totaling \NphotTot\ photometric data points (\NphotTotDet\ robust transient detections) in $grizy$ filters for \NIItotal\ individual SNe (\NIIPtotalF\ SNe~IIP).  We have developed a Bayesian light curve fitting methodology for SNe~II based on a physically-motivated 5-component segmentation of the SN~IIP light curve (Section~\ref{ssec:mod}).  We present an implementation of our SN~IIP model for use with the Hamiltonian Monte Carlo library \textit{Stan} in Appendix~\ref{ap:stan}.  We have interpreted our light curve modeling in terms of the hydrodynamic Type~IIP SN progenitor model grid of \cite{Kasen09}.

The primary conclusions of this work are as follows:
    
\begin{itemize}
\item We present photometric $K$-corrections and SN~IIP light curve templates in the $grizy$ bands (Sections~\ref{ssec:Kcor} and \ref{sec:res:stack}).  Our templates are based on \stackNtemptotalALL\ individual photometric detections for \stackNacceptr\ individual SNe~IIP from $-15$ to $+114$ days from peak magnitude.
\item Consistent with theoretical expectations, our SN~IIP sample spans a diverse range of light curve parameters: a factor of $\sim20$ in plateau phase decay rate (Figure~\ref{fig:betakde}), $\sim2$ orders of magnitude in ejected $^{56}$Ni (Figure~\ref{fig:MNidist}), a factor of $\sim2$ in plateau duration (Figure~\ref{fig:tpfdist}), and $\gtrsim4$~mag in absolute magnitude (Figure~\ref{fig:mpeakdist}).  This evidence stands in contrast to recent suggestions by \cite{Arcavi12} and \cite{Poznanski13} that the SN~IIP plateau duration distribution, a critical observational parameter tied to progenitor initial mass, is tightly distributed (Section~\ref{sec:res:tp}).
\item Addressing a longstanding debate in the literature, we have searched for the existence of a fast declining ``SN~IIL'' sub-population in the decline rate distribution of our SN~IIP sample (Section~\ref{sec:res:IIL}).  We find no evidence for a discontinuity in this distribution for any photometric band, questioning the existence of a discrete SN~IIL sub-population.
\item We identify a highly significant statistical correlation between the peak magnitude and plateau phase decline rate of SNe~IIP (Section~\ref{sec:res:pcor}).  Together with the previous results, this  supports the interpretation of core-collapse among hydrogen-rich red supergiants as a predominantly single parameter family of explosions, whose observational behavior is determined primarily by the explosion energy and likely set by the initial mass of the progenitor star.  This represents an independent discovery and confirmation of results recently reported in \cite{Anderson14phot}.
\item Through the largest systematic comparison to date of SN~IIP lightcurves to hydrodynamic progenitor models, we have derived mass, radius, and explosion energy estimates for the objects in our sample (Section~\ref{sec:prog:mass}).  However, we find that the available theoretical model grids are insufficient to cover the full range of observed variation in SN~IIP light curve properties.  We point to the need for additional hydrodynamic modeling to produce updated and expanded self-consistent model grids, particularly in the low luminosity regime.
\item Though our progenitor inferences are based on hydrodynamic light curve models, which are known to produce systematically higher masses than direct progenitor detection searches, we do not find evidence for an absence of high-mass SN~IIP progenitors (Section~\ref{sec:prog:RSGprob}).  We point to future direct progenitor detections of luminous SNe~IIP as having the potential to ease the discrepancy between the maximum SN~IIP progenitor mass identified by various theoretical and observational methods, known as the Red Supergiant Problem.
\end{itemize}

In a companion paper, \cite{Sanders14Unsup}, we discuss and demonstrate a hierarchical expansion of the model presented here to provide a general framework for the analysis of supernova light curve populations in the coming era of next generation wide field transient searches.  We advocate continued investment in statistical and computational tools in the future as a means to compensate for the relative decline anticipated in the availability of detailed spectroscopic and other follow-up information on individual transients.

\acknowledgements
\label{sec:ackn}

We thank K. Mandel for sage guidance and many helpful conversations; D. Kasen for his thoughtful comments on this work; an anonymous referee for their very helpful review; and the \textit{Stan} team for their excellent modeling language and HMC sampler.

The Pan-STARRS1 Surveys (PS1) have been made possible through contributions of the Institute for Astronomy, the University of Hawaii, the Pan-STARRS Project Office, the Max-Planck Society and its participating institutes, the Max Planck Institute for Astronomy, Heidelberg and the Max Planck Institute for Extraterrestrial Physics, Garching, The Johns Hopkins University, Durham University, the University of Edinburgh, Queen's University Belfast, the Harvard-Smithsonian Center for Astrophysics, the Las Cumbres Observatory Global Telescope Network Incorporated, the National Central University of Taiwan, the Space Telescope Science Institute, the National Aeronautics and Space Administration under Grant No. NNX08AR22G issued through the Planetary Science Division of the NASA Science Mission Directorate, the National Science Foundation under Grant No. AST-1238877, the University of Maryland, and Eotvos Lorand University (ELTE).

Observations reported here were obtained at the MMT Observatory, a joint facility of the Smithsonian Institution and the University of Arizona.  This paper includes data gathered with the 6.5 meter Magellan Telescopes located at Las Campanas Observatory, Chile.  Based on observations obtained at the Gemini Observatory, which is operated by the Association of Universities for Research in Astronomy, Inc., under a cooperative agreement with the NSF on behalf of the Gemini partnership: the National Science Foundation (United States), the National Research Council (Canada), CONICYT (Chile), the Australian Research Council (Australia), Minist\'{e}rio da Ci\^{e}ncia, Tecnologia e Inova\c{c}\~{a}o (Brazil) and Ministerio de Ciencia, Tecnolog\'{i}a e Innovaci\'{o}n Productiva (Argentina).  The data presented here were obtained in part with ALFOSC, which is provided by the Instituto de Astrofisica de Andalucia (IAA) under a joint agreement with the University of Copenhagen and NOTSA.

Support for this work was provided by the David and Lucile Packard Foundation Fellowship for Science and Engineering awarded to A.M.S.  M.B. is supported under EPSRC grant EP/J016934/1

{\it Facilities:} \facility{PS1}  \facility{MMT} \facility{Magellan:Baade} \facility{Magellan:Clay} \facility{Gemini} \facility{NOT} 

\bibliographystyle{fapj}

\clearpage

\begin{appendix}
\section{Appendix A: Individual Light Curve Stan Model}
\label{ap:stan}

Below we reproduce the full Bayesian model for our 5-component segmented SN~II light curve model described in Section~\ref{sec:mod:fit}, in the \textit{Stan} modeling language.  The Stan model specification format is documented in the Stan Modeling Language Users Guide and Reference Manual \citep{stan-manual:2014}.  

The model takes the following data as input: \texttt{N\_obs}, the total number of photometric data points; \texttt{N\_filt}, the number of photometric filters; \texttt{t}, a vector of MJD dates of the photometric observations; \texttt{fL}, a vector of luminosities corresponding to the photometric observations (with units as described in Section~\ref{ssec:mod}); \texttt{dfL} a corresponding vector of luminosity uncertainties; \texttt{z} the redshift; \texttt{t0\_mean} an initial estimate of the explosion date (for initialization and for centering the explosion date prior distribution); \texttt{J} a vector of integers specifying the filter ID of each photometric observation; \texttt{Kcor\_N}, a matrix of pre-computed $K$-corrections for each filter, in magnitudes with spacing of 1~day; \texttt{fluxscale} the zero-point of the luminosity unit system ($\texttt{fluxscale} = 10^7$ in the system we have employed); and \texttt{duringseason}, a boolean value specifying whether the object exploded within or between observing seasons, for selection of the explosion date prior distribution parameters.  The calculation of the model light curve flux and application of the $K$-correction values is performed in the \texttt{transformed parameters} section, and the prior and likelihood calculations are performed in the \texttt{model} section.  Certain vector-valued prior distribution parameters are specified in the \texttt{transformed data} section for convenience.

The \textit{Stan} model is then compiled and run \citep{stan-manual:2014} to yield MCMC samples from the posterior distribution of light curve parameters.  We have used PyStan version 2.2.0\footnote{\url{https://github.com/stan-dev/stan/releases/tag/v2.2.0}}.

We configured the \textit{No-U-Turn Sampler} to use fixed $0$ initialization of the parameter values, an adaptation phase of 250~steps, a maximum treedepth of 22, and otherwise employed the default sampler parameters.  Using this configuration, we achieve a Gelman-Rubin between-to-within chain variance ratio of $\hat{R}<1.02$ for 95\% of parameters, indicating excellent convergence.  The typical run time is 1~minute per chain.

{\footnotesize
\begin{verbatim}
data {
  int<lower=0> N_obs;                             
  int<lower=0> N_filt;                            
  vector[N_obs] t;                                
  vector[N_obs] fL;                               
  vector[N_obs] dfL;                              
  real z;                                 
  real t0_mean;                           
  int<lower=1,upper=N_filt> J[N_obs];             
  int<lower=0> Kcor_N;                            
  real Kcor[N_filt,Kcor_N];                 
  real<lower=0> fluxscale;                        
  real<lower=0,upper=1> duringseason;     
}
transformed data {
    vector[N_filt] prior_tp;
    vector[N_filt] prior_sig_tp;
    vector[N_filt] prior_lbeta1;
    vector[N_filt] prior_sig_lbeta1;
    vector[N_filt] prior_lbeta2;
    vector[N_filt] prior_sig_lbeta2;
    prior_tp[1] <- log(5); 
        prior_tp[2] <- log(8); 
        prior_tp[3] <- log(14); 
        prior_tp[4] <- log(20); 
        prior_tp[5] <- log(30);
    prior_sig_tp[1] <- 0.3; 
        prior_sig_tp[2] <- 0.3; 
        prior_sig_tp[3] <- 0.3; 
        prior_sig_tp[4] <- 0.3; 
        prior_sig_tp[5] <- 0.3;
    prior_lbeta1[1] <- -2.1; 
        prior_lbeta1[2] <- -2.3; 
        prior_lbeta1[3] <- -3.3; 
        prior_lbeta1[4] <- -3.8; 
        prior_lbeta1[5] <- -4.0;
    prior_sig_lbeta1[1] <- 0.6; 
        prior_sig_lbeta1[2] <- 0.8; 
        prior_sig_lbeta1[3] <- 1.2; 
        prior_sig_lbeta1[4] <- 1.5; 
        prior_sig_lbeta1[5] <- 2;
    prior_lbeta2[1] <- -3.4; 
        prior_lbeta2[2] <- -4; 
        prior_lbeta2[3] <- -4.1; 
        prior_lbeta2[4] <- -4.4; 
        prior_lbeta2[5] <- -4.9;
    prior_sig_lbeta2[1] <- 1; 
        prior_sig_lbeta2[2] <- 1.2; 
        prior_sig_lbeta2[3] <- 1.2; 
        prior_sig_lbeta2[4] <- 1.5; 
        prior_sig_lbeta2[5] <- 1.5;
}
parameters {  
    real pt0;
    vector<lower=0>[N_filt] t1;
    vector<lower=0>[N_filt] t2;
    vector<lower=0>[N_filt] td;
    vector<lower=0>[N_filt] tp;
    vector<upper=0>[N_filt] lalpha; 
    vector<upper=0>[N_filt] lbeta1;
    vector<upper=0>[N_filt] lbeta2;
    vector<upper=0>[N_filt] lbetadN;
    vector<upper=0>[N_filt] lbetadC; 
    vector<lower=0>[N_filt] Mp; 
    vector[N_filt] Yb;
    vector<lower=0>[N_filt] V;
}
transformed parameters {
    vector[N_obs] mm;                         
    vector[N_obs] dm;                         
    vector<lower=0>[N_filt] M1;
    vector<lower=0>[N_filt] M2;
    vector<lower=0>[N_filt] Md;
    M1 <- Mp ./ exp( exp(lbeta1) .* tp );
    M2 <- Mp .* exp( -exp(lbeta2) .* t2 );
    Md <- M2 .* exp( -exp(lbetadN) .* td );
    for (n in 1:N_obs) {      
        real N_SNc;                                      
        int Kc_up;                                       
        int Kc_down;                                     
        real t_exp;                                      
        int j;                                           
        int k;                                           
        real mm_1;
        real mm_2;
        real mm_3;
        real mm_4;
        real mm_5;
        real mm_6;
        j <- J[n];
        t_exp <- ( t[n] - (t0_mean + pt0) ) / (1 + z);
        if (t_exp<0) {                                                                      
            mm_1 <- Yb[j];
        } else {
            mm_1 <- 0;
        }
        if ((t_exp>=0) && (t_exp < t1[j])) {                                                                      
            mm_2 <- Yb[j] + M1[j] * pow(t_exp / t1[j] , exp(lalpha[j]));
        } else {
            mm_2 <- 0;
        }
        if ((t_exp >= t1[j]) && (t_exp < t1[j] + tp[j])) {                                                                      
            mm_3 <- Yb[j] + M1[j] * exp(exp(lbeta1[j]) * (t_exp - t1[j]));
        } else {
            mm_3 <- 0;
        }
        if ((t_exp >= t1[j] + tp[j]) && (t_exp < t1[j] + tp[j] + t2[j])) {                                                                      
            mm_4 <- Yb[j] + Mp[j] * exp(-exp(lbeta2[j]) * (t_exp - t1[j] - tp[j]));
        } else {
            mm_4 <- 0;
        }
        if ((t_exp >= t1[j] + tp[j] + t2[j]) && (t_exp < t1[j] + tp[j] + t2[j] + td[j])) {                                                                      
            mm_5 <- Yb[j] + M2[j] * exp(-exp(lbetadN[j]) * (t_exp - t1[j] - tp[j] - t2[j]));
        } else {
            mm_5 <- 0;
        }
        if (t_exp >= t1[j] + tp[j] + t2[j] + td[j]) {                                                                      
            mm_6 <- Yb[j] + Md[j] * exp(-exp(lbetadC[j]) * (t_exp - t1[j] - tp[j] - t2[j] - td[j]));
        } else {
            mm_6 <- 0;
        }
        dm[n] <- sqrt(pow(dfL[n],2) + pow(V[j],2));
        if (t_exp<0) {
            N_SNc <- 0;
        } else if  (t_exp<Kcor_N-2){                    
            Kc_down <- 0;
            while ((Kc_down+1) < t_exp) {               
                Kc_down <- Kc_down + 1; 
            }
            Kc_up <- Kc_down+1;
            N_SNc <- Kcor[j,Kc_down+1] + (t_exp - floor(t_exp)) * (Kcor[j,Kc_up+1]-Kcor[j,Kc_down+1]);
        } else {                                        
            N_SNc <- Kcor[j,Kcor_N];
        }
        mm[n] <- (mm_1+mm_2+mm_3+mm_4+mm_5+mm_6) 
                 / (pow(10, N_SNc/(-2.5)));
    }
}
model {
    if (duringseason == 1) {
        pt0 ~ skew_normal(-1, 1, -0.5);
    } else {
        pt0 ~ skew_normal(-30, 20, -1);
    }
    t1 ~ lognormal(log(1), 0.3);
    tp ~ lognormal(prior_tp, prior_sig_tp);
    t2 ~ lognormal(log(100), 0.3);
    td ~ lognormal(log(10), 0.5);
    lalpha ~ normal(-1, 0.3); 
    lbeta1 ~ normal(prior_lbeta1, prior_sig_lbeta1);
    lbeta2 ~ normal(prior_lbeta2, prior_sig_lbeta2);
    lbetadN ~ normal(-3, 0.5);
    lbetadC ~ normal(-5, 1);
    Mp ~ lognormal(log(1), 0.7); 
    Yb ~ normal(0, 0.3);
    V ~ cauchy(0, 0.01);
    fL ~ normal(mm,dm);
}
\end{verbatim}
}

\section{Appendix B: Linear Relation With Intrinsic Scatter Stan Model}
\label{ap:stan2}

Below we reproduce the full probabilistic model describing a linear relation between bivariate data with intrinsic scatter, which we apply to the SN~IIP Decline rate--peak magnitude in Section~\ref{sec:res:pcor}, specified in the \textit{Stan} modeling language and following \citep{HoggModel}.  

The model takes the following data as input: \texttt{N}, the number of bivariate observations; \texttt{x} and \texttt{y}, the mean observed values of the first and second covariates; \texttt{dx} and \texttt{dy}, the uncertainty (standard deviation) of each observation projected along the axis of each covariate; and \texttt{dxy} the covariance of each observation.  In the \texttt{transformed parameters} section, the sampled angle of the linear relation from the $x$-axis (\texttt{theta}) and its perpendicular distance (\texttt{b\_perp}) from the origin are used to calculate the perpendicular distance of each observation from the ridgeline of the sampled relation and its projected variance.  The prior and likelihood calculations are performed in the \texttt{model} section, accounting for both the projected observational variance and the intrinsic variance (\texttt{V}).  The \texttt{b\_perp} and \texttt{theta} parameters are transformed into traditional linear slope and offset parameters (\texttt{m,b}) in the \texttt{generated quantities} section for convenience.

The \textit{Stan} model is then compiled and run as described in Appendix~\ref{ap:stan}.

{\footnotesize
\begin{verbatim}
data {
  int<lower=0> N; 
  vector[N] x; 
  vector[N] y; 
  vector[N] dx; 
  vector[N] dy; 
  vector[N] dxy; 
}

parameters {
  real<lower=-pi()/2.,upper=pi()/2.> theta; 
  real b_perp;  
  real<lower=0> V;
}

transformed parameters {
  unit_vector[2] v;   
  vector[N] Delta;  
  matrix[N,2] Z;    
  cov_matrix[2] S[N];  
  vector[N] Sigma_squared;  
  
  v[1] <- -sin(theta);
  v[2] <- cos(theta);
  for (n in 1:N) {
    Z[n,1] <- x[n];
    Z[n,2] <- y[n];
    S[n][1,1] <- pow(dx[n],2);
    S[n][2,2] <- pow(dy[n],2);
    S[n][1,2] <- dxy[n];
    S[n][2,1] <- S[n][1,2];
    Sigma_squared[n] <- v' * S[n] * v;
    Delta[n] <- v' * Z[n]' - b_perp;
  }
}

model {
  V ~ cauchy(0,2.5);
  for (n in 1:N) {
    0 ~ normal(Delta[n],sqrt(Sigma_squared[n] + V));
  }
}

generated quantities {
  real b;                             
  real m;            
  
  m <- tan(theta);
  b <- b_perp / cos(theta);
}\end{verbatim}
}

\end{appendix}

\clearpage
\LongTables
\begin{deluxetable*}{lllllllllll}
\tablecaption{Pan-STARRS1 Type~IIP SN Sample\label{tab:obspars}}
\tabletypesize{\scriptsize}
\tablehead{\colhead{PS1 SN} & \colhead{Field} & \colhead{RA} & \colhead{DEC} & \colhead{Disc. MJD} & \colhead{$z$} & \colhead{$N_g$} & \colhead{$N_r$} & \colhead{$N_i$} & \colhead{$N_z$} & \colhead{$N_y$}}
\startdata
PS1--10a & 02 & 03:29:48.735 & --29:04:29.91 & 55207 & 0.071 & 2 & 2 & 5 & 4 & 0\\ 
PS1--10b & 03 & 08:37:02.587 & +43:44:16.85 & 55207 & 0.260 & 4 & 4 & 5 & 0 & 0\\ 
PS1--10q & 06 & 12:19:37.533 & +46:01:56.03 & 55212 & 0.057 & 15 & 17 & 16 & 13 & 0\\ 
PS1--10t & 04 & 09:55:30.959 & +01:29:01.18 & 55214 & 0.093 & 11 & 11 & 11 & 11 & 0\\ 
PS1--10ae & 04 & 10:02:09.742 & +01:14:00.93 & 55242 & 0.086 & 4 & 5 & 8 & 6 & 0\\ 
PS1--10lv & 09 & 22:13:00.969 & +00:46:24.49 & 55389 & 0.127 & 10 & 21 & 22 & 23 & 0\\ 
PS1--10mb & 08 & 16:11:44.435 & +56:05:08.39 & 55389 & 0.064 & 2 & 12 & 13 & 10 & 0\\ 
PS1--10zu & 01 & 02:18:54.489 & --04:14:42.96 & 55407 & 0.059 & 11 & 18 & 21 & 24 & 0\\
\enddata
\tablecomments{This table includes only SNe classified as SNe~IIP by the criteria of Section~\ref{sec:mod:class}.  PS1~SN is the PS1 designation of the transient; field is the PS1 Medium Deep Field where the transient was discovered, RA and DEC are in J2000 coordinates, Disc. MJD is the date of discovery, $z$ is the redshift obtained from our follow-up spectroscopy (Section~\ref{sec:obs:spec}), and $N_x$ is the number of detections of the object included in our dataset in each PS1 filter (additional photometric non-detections are also used to constrain the explosion date and decay behavior).}
\end{deluxetable*}

\clearpage
\setlength{\tabcolsep}{2pt}
\LongTables
\begin{deluxetable*}{lrrrrrrrrrrrr}
\tabletypesize{\scriptsize}
\tablecaption{Pan-STARRS1 Type~IIP SN Light Curve Parameters\label{tab:lcpars}}
\tablehead{\colhead{SN} & \colhead{$t_0$} & \colhead{$\log \alpha$} & \colhead{$\log \beta_1$} & \colhead{$\log \beta_2$} & \colhead{$\log \beta_{dN}$} & \colhead{$\log \beta_{dC}$} & \colhead{$t_1$} & \colhead{$t_p$} & \colhead{$t_2$} & \colhead{$t_d$} & \colhead{$M_p$} & \colhead{$V$}}  
\startdata
\multicolumn{13}{l}{$g$-band} \\ 
PS1--10a & $55200.9^{+0.8}_{-0.8}$ & $-1.0^{+0.3}_{-0.3}$ & $-2.4^{+0.5}_{-0.5}$ & $-3.1^{+0.3}_{-0.4}$ & $-2.9^{+0.5}_{-0.5}$ & $-5.0^{+1.2}_{-1.1}$ & $1.0^{+0.3}_{-0.2}$ & $5^{+1}_{-1}$ & $106^{+33}_{-25}$ & $10^{+7}_{-4}$ & $2.10^{+0.45}_{-0.25}$ & $0.005^{+0.006}_{-0.003}$ \\ 
PS1--10b & $55204.0^{+0.4}_{-0.5}$ & $-1.0^{+0.3}_{-0.3}$ & $-2.8^{+0.3}_{-0.3}$ & $-2.7^{+0.2}_{-0.2}$ & $-3.0^{+0.5}_{-0.5}$ & $-4.9^{+1.0}_{-1.1}$ & $1.0^{+0.4}_{-0.3}$ & $7^{+1}_{-1}$ & $102^{+38}_{-28}$ & $10^{+6}_{-4}$ & $1.93^{+0.11}_{-0.10}$ & $0.008^{+0.017}_{-0.006}$ \\ 
PS1--10q & $55200.4^{+0.7}_{-0.7}$ & $-1.0^{+0.3}_{-0.3}$ & $-2.4^{+0.5}_{-0.5}$ & $-3.1^{+0.0}_{-0.0}$ & $-2.9^{+0.5}_{-0.6}$ & $-4.9^{+1.0}_{-1.0}$ & $1.0^{+0.4}_{-0.3}$ & $4^{+1}_{-1}$ & $109^{+31}_{-24}$ & $10^{+8}_{-4}$ & $0.77^{+0.03}_{-0.03}$ & $0.016^{+0.004}_{-0.003}$ \\ 
PS1--10t & $55205.2^{+0.9}_{-0.8}$ & $-1.0^{+0.3}_{-0.3}$ & $-2.2^{+0.6}_{-0.5}$ & $-3.5^{+0.1}_{-0.1}$ & $-3.1^{+0.5}_{-0.4}$ & $-5.1^{+1.1}_{-0.9}$ & $1.0^{+0.3}_{-0.3}$ & $5^{+1}_{-1}$ & $97^{+32}_{-24}$ & $10^{+7}_{-4}$ & $0.37^{+0.02}_{-0.02}$ & $0.011^{+0.007}_{-0.007}$ \\ 
PS1--10ae & $55239.3^{+0.1}_{-0.1}$ & $-1.1^{+0.3}_{-0.3}$ & $-2.7^{+0.5}_{-0.5}$ & $-3.3^{+0.1}_{-0.1}$ & $-3.0^{+0.5}_{-0.5}$ & $-5.6^{+0.7}_{-0.7}$ & $0.9^{+0.2}_{-0.2}$ & $3^{+1}_{-0}$ & $76^{+24}_{-22}$ & $10^{+5}_{-4}$ & $0.84^{+0.02}_{-0.02}$ & $0.005^{+0.006}_{-0.003}$ \\ 
PS1--10lv & $55379.1^{+3.0}_{-3.9}$ & $-1.0^{+0.3}_{-0.3}$ & $-2.1^{+0.6}_{-0.6}$ & $-3.1^{+0.0}_{-0.0}$ & $-3.0^{+0.5}_{-0.5}$ & $-4.9^{+0.9}_{-0.8}$ & $1.0^{+0.3}_{-0.3}$ & $5^{+2}_{-1}$ & $99^{+34}_{-23}$ & $10^{+6}_{-4}$ & $1.96^{+0.25}_{-0.16}$ & $0.007^{+0.011}_{-0.005}$ \\
\enddata
\tablecomments{Light curve parameters are listed separately for each photometric band, in sequence.  The parameters are defined in Section~\ref{ssec:mod}.  This table includes only SNe classified as SNe~IIP by the criteria of Section~\ref{sec:mod:class}.}
\end{deluxetable*}
\setlength{\tabcolsep}{6pt}

\clearpage
\begin{deluxetable*}{lrrrrr}
\tablecaption{SN~IIP $grizy$ Template Light Curves\label{tab:templ}}
\tabletypesize{\scriptsize}
\tablehead{\colhead{Epoch} & \colhead{$g$} & \colhead{$r$} & \colhead{$i$} & \colhead{$z$} & \colhead{$y$}}
\startdata
-15 & $\ldots$ & $\ldots$ & $\ldots$ & $\ldots$ & $\ldots$\\ 
-12 & $\ldots$ & $\ldots$ & $\ldots$ & $1.32^{+0.29}_{-0.33}$ & $\ldots$\\ 
-9 & $0.32^{+0.14}_{-0.14}$ & $0.61^{+0.38}_{-0.26}$ & $0.50^{+1.28}_{-0.19}$ & $0.88^{+0.59}_{-0.34}$ & $0.26^{+0.43}_{-0.38}$\\ 
-6 & $0.21^{+0.22}_{-0.14}$ & $0.43^{+0.33}_{-0.29}$ & $0.34^{+1.00}_{-0.16}$ & $0.65^{+0.55}_{-0.35}$ & $0.31^{+0.23}_{-0.45}$\\ 
-3 & $0.09^{+0.22}_{-0.09}$ & $0.15^{+0.46}_{-0.15}$ & $0.25^{+0.60}_{-0.22}$ & $0.51^{+0.38}_{-0.40}$ & $0.18^{+0.20}_{-0.29}$\\ 
0 & $0.10^{+0.13}_{-0.07}$ & $0.07^{+0.24}_{-0.08}$ & $0.08^{+0.24}_{-0.10}$ & $0.34^{+0.37}_{-0.30}$ & $0.12^{+0.20}_{-0.17}$\\ 
3 & $0.14^{+0.13}_{-0.12}$ & $0.06^{+0.07}_{-0.07}$ & $0.04^{+0.13}_{-0.10}$ & $0.25^{+0.30}_{-0.23}$ & $0.09^{+0.19}_{-0.16}$\\ 
6 & $0.22^{+0.24}_{-0.14}$ & $0.09^{+0.12}_{-0.08}$ & $0.05^{+0.13}_{-0.11}$ & $0.16^{+0.33}_{-0.14}$ & $0.08^{+0.16}_{-0.22}$\\
\enddata
\tablecomments{The reported epoch is in rest frame days since peak.  The $grizy$ photometry is reported in terms of magnitude below peak; the ranges reflect the 16-84th percentile variation in observed light curve behavior.}
\end{deluxetable*}

\clearpage
\LongTables
\begin{deluxetable*}{lllllllll}
\tablecaption{Light curve and modeled progenitor parameters for PS1 SNe~IIP\label{tab:ppars}}
\tabletypesize{\scriptsize}
\tablehead{\colhead{PS1 SN} & \colhead{$M_{p,r}$} & \colhead{$t_{plateau,r}$} & \colhead{$\log_{10}(L_{50})$} & \colhead{$\Delta m_{15,r}$} & \colhead{$M_{\rm Ni}$} & \colhead{$M_{in}$} & \colhead{$R_0$} & \colhead{$E_{51}$}\\ 
 & \colhead{(mag)} & \colhead{(days)} & \colhead{(ergs~s$^{-1}$)} & \colhead{(mag)} & \colhead{(mag)} & \colhead{($M_\odot$)} & \colhead{($R_\odot$)} & \colhead{($10^{51}$ ergs)}}
\startdata
PS1--10a & $-18.11^{+0.17}_{-0.20}$ & $114^{+32}_{-30}$ & $42.63^{+0.55}_{-0.41}$ & $-0.52^{+0.28}_{-0.38}$ & $0.15^{+0.32}_{-0.11}$ & \nodata & \nodata & \nodata\\ 
PS1--10q & $-17.06^{+0.04}_{-0.05}$ & $96^{+4}_{-2}$ & $42.42^{+0.15}_{-0.12}$ & $-0.26^{+0.02}_{-0.02}$ & $0.11^{+0.19}_{-0.07}$ & \nodata & \nodata & \nodata\\ 
PS1--10ae & $-17.17^{+0.04}_{-0.04}$ & $112^{+28}_{-26}$ & $41.93^{+0.20}_{-0.13}$ & $-0.30^{+0.03}_{-0.04}$ & $0.06^{+0.07}_{-0.04}$ & \nodata & \nodata & \nodata\\ 
PS1--10lv & $-18.43^{+0.04}_{-0.03}$ & $109^{+21}_{-7}$ & $43.12^{+0.04}_{-0.06}$ & $-0.27^{+0.01}_{-0.01}$ & $0.20^{+0.14}_{-0.14}$ & \nodata & \nodata & \nodata\\ 
PS1--10mb & $-15.46^{+0.15}_{-0.17}$ & $103^{+27}_{-20}$ & $41.80^{+0.20}_{-0.15}$ & $-0.08^{+0.04}_{-0.05}$ & $0.06^{+0.06}_{-0.04}$ & \nodata & \nodata & \nodata\\ 
PS1--10zu & $-16.81^{+0.14}_{-0.12}$ & $110^{+5}_{-26}$ & $42.30^{+0.10}_{-0.14}$ & $-0.10^{+0.02}_{-0.02}$ & $0.07^{+0.04}_{-0.04}$ & \nodata & \nodata & \nodata\\ 
PS1--10abd & $-18.55^{+0.11}_{-0.11}$ & $92^{+38}_{-9}$ & $42.64^{+0.11}_{-0.14}$ & $-0.33^{+0.02}_{-0.02}$ & $0.09^{+0.09}_{-0.05}$ & \nodata & \nodata & \nodata\\ 
PS1--10agh & $-18.57^{+0.05}_{-0.05}$ & $100^{+18}_{-15}$ & $42.93^{+0.04}_{-0.04}$ & $-0.36^{+0.01}_{-0.01}$ & $0.11^{+0.16}_{-0.07}$ & $12.8^{+7.0}_{-4.2}$ & $670^{+370}_{-210}$ & $6.53^{+3.61}_{-2.14}$\\
\enddata
\tablecomments{The $r$-band peak magnitude ($M_{p,r}$) is derived directly from the model fits (Section~\ref{sec:mod:fit}), including $K$-corrections, but not corrected for extinction.  The $r$-band plateau duration ($t_{plateau,r}$) is in rest frame days, not corrected for contamination by $^{56}$Ni.  The pseudo-bolometric (OIR) luminosity at 50~days ($L_{50}$) is measured from the fits to the PS1 photometry; the value quoted does not include the bolometric correction described in Section~\ref{sec:mod:fit}.  The decline rate in $r$-band ($\Delta m_{15,r}$; rest frame days) is measured from the same model fits.  The ejected nickel mass is estimated from comparison of the late-time bolometric light curve to that of SN~1987A (but see Section~\ref{sec:res:decay} for a discussion of reliability).  The progenitor initial mass ($M_{in}$), radius ($R_0$), and explosion energy ($E_{51}$) estimates come from the comparisons (see Section~\ref{sec:res:prog}) to the models of \cite{Kasen09}; but see Section~\ref{sec:prog:mass} for a discussion of reliability.  These theoretical estimates are reported only if the central mass estimate falls within the model grid ($12-25~M_\odot$); but the full posterior distribution for the reported values includes extrapolations to values beyond the model grid, as reflected in the reported uncertainties.  All reported values represent [16,50,84]th percentiles of the posterior distribution.}
\end{deluxetable*}

\end{document}